\documentclass[aps, reprint, amsmath,amssymb, prl, superscriptaddress]{revtex4-2}
\usepackage{svg}
\usepackage{color}
\usepackage{graphicx}
\usepackage{dcolumn}
\usepackage{bm}
\usepackage{physics}
\usepackage{hyperref}
\usepackage{url}
\usepackage{amsmath}
\usepackage{tikz-cd}
\usepackage{notes2bib}
\usepackage{ dsfont }
\usepackage{svg}
\usepackage{verbatim}
\usepackage{amssymb}
\usepackage{amsfonts}              
\usepackage{amsthm}                
\usepackage{mathtools}
\usepackage{qcircuit}
\usepackage[normalem]{ulem}
\usepackage{xcolor}

\usepackage{enumitem}

\DeclarePairedDelimiterX{\barpair}[2]{(}{)}{%
  #1\;\delimsize\|\;#2%
}
\theoremstyle{definition}
\newtheorem{theorem}{Theorem}
\newtheorem{prop}[theorem]{Proposition}

\newtheorem{lemma}[theorem]{Lemma}
\newtheorem{definition}[theorem]{Definition}
\newtheorem{corollary}[theorem]{Corollary}

\newtheorem{stheorem}{S.Theorem}

\newtheorem{slemma}[stheorem]{S.Lemma}

\newtheorem{scorollary}[stheorem]{S.Corollary}

\newcommand{\Ad}{\text{Ad}}
\newcommand{\id}{\text{id}}
\newcommand{\mf}[1]{\mathfrak{#1}}
\newcommand{\mcal}[1]{\mathcal{#1}}

\newcommand{\jami}{Jamio{\l}kowski }

\newcommand{\mds}[1]{\mathds{#1}}
\newcommand{\supp}[1]{\text{supp}\left(#1\right)}

\usepackage[normalem]{ulem}
\bibnotesetup{note-name    = ,use-sort-key = false}
\begin{document}

\title{Catalysis always degrades external quantum correlations}

\author{Seok Hyung Lie}
\affiliation{%
 School of Physical and Mathematical Sciences, Nanyang Technological University, 21 Nanyang Link, Singapore, 637371
}%
\author{Nelly H.Y. Ng}
\affiliation{%
 School of Physical and Mathematical Sciences, Nanyang Technological University, 21 Nanyang Link, Singapore, 637371
}%

\date{\today}

\begin{abstract}
Catalysts used in quantum resource theories need not be in isolation and therefore are possibly correlated with external systems, which the agent does not have access to. Do such correlations help or hinder catalysis, and does the classicality or quantumness of such correlations matter? To answer this question, we first focus on the existence of a non-invasively measurable observable that yields the same outcomes for repeated measurements, since this signifies macro-realism, a key property distinguishing classical systems from quantum systems. We show that a system quantumly correlated with an external system so that the joint state is necessarily perturbed by any repeatable quantum measurement, also has the same property against general quantum channels. Our full characterization of such systems called \emph{totally quantum systems}, solves the open problem of characterizing tomographically sensitive systems raised in [Lie and Jeong, Phys. Rev. Lett. 130, 020802 (2023)]. An immediate consequence is that a totally quantum system cannot catalyze any quantum process, even when a measure of correlation with its environment is arbitrarily low. It generalizes to a stronger result, that the mutual information of totally quantum systems cannot be used as a catalyst either. These results culminate in the conclusion that, out of the correlations that a generic quantum catalyst has with its environment, only classical correlations allow for catalysis, and therefore using a correlated catalyst is equivalent to using an ensemble of uncorrelated catalysts. 
\end{abstract}

\pacs{Valid PACS appear here}
\maketitle
\textit{Introduction.}---  Catalysis in quantum resource theory, a concept inspired by catalysis in chemistry, is a paradigm that utilizes some quantum resources without altering or deteriorating it, while expanding the set of accessible quantum states (or channel) transformations~\cite{datta2022catalysis}. 
Initially, catalysis was often studied under the condition of being uncorrelated with the final state of the system~\cite{jonathan1999entanglement,daftuar2001mathematical,klimesh2007inequalities,aubrun2008catalytic,campbell2011catalysis,brandao2015second}.
In recent years, however, a notable trend is the investigation of how the power of catalysis, in enabling state transitions, can be increased by allowing correlations to persist between system and catalyst after the process \cite{aaberg2014catalytic,muller2018correlating,shiraishi2021quantum,wilming2017axiomatic,boes2019neumann,rethinasamy2020relative,kondra2021catalytic,lie2020randomness,lie2020uniform,lie2021correlational,lie2022delocalized}. The relaxation of this constraint simplifies the conditions for state transition significantly, and often leads to a characterization of von Neumann quantities (e.g. entropy) -- which have a strong operational significance previously only in the asymptotic i.i.d. regime -- in one-shot settings. It is argued that in certain scenarios, the resultant correlation can be ignored, which assumes that only the marginal state of catalyst is relevant when catalyst is separated from the system.

\begin{figure}[t]
    \includegraphics[width=.45\textwidth]{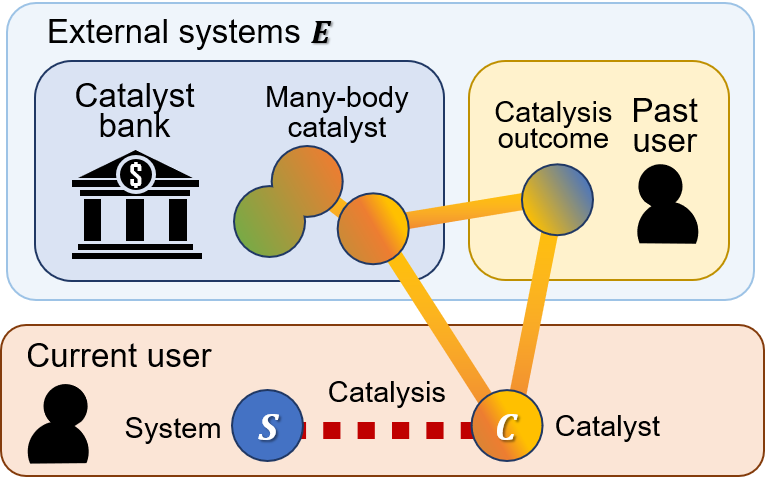}
    \caption{A quantum resource catalyst could be correlated with external systems inaccessible to a user in many plausible scenarios. First, a catalyst can be a part of multipartite collection of catalysts of the bank. In this case, it is natural for the bank to demand the multipartite state to remain intact after each catalysis. Second, even for an initially uncorrelated catalyst, after a round of correlation-forming catalysis, it remains correlated with its previous user. The same user can borrow the catalyst again, and it is natural to expect the relation with the catalyst to remain the same as the previous round of catalysis.}\label{fig:TQcatalysis}
\end{figure}

However, this line of thought clashes with an often-used concept of `catalyst bank' \cite{sparaciari2018multi,feng2005catalyst,fritz2017resource}, a hypothetical entity that lends quantum resource catalysts to (possibly many) agents and retrieves thereafter (See FIG. \ref{fig:TQcatalysis}.) A catalyst could be only a part of a large collection of quantum systems possessed by the bank. In this case, it is operationally natural for the bank to require the agent who borrows the catalyst to return it in a fashion that the whole quantum system stays in the same state, even though the agent used only a small portion. Even if the catalyst is prepared uncorrelated with other systems, after a single round of correlated catalysis by some user, the catalyst will form correlation. When the next user borrows the catalyst, again, it is natural to require the correlations to be preserved, as nothing forbids the same user to borrow the same catalyst twice, and not wasting any resource not possessed by oneself is the prime premise of resource catalysis.

Some studies on correlating catalysis deal with such potential problems, by showing that the amount of correlation formed in catalysis can be made arbitrarily small, e.g. \cite{muller2018correlating, kondra2021catalytic}. However, we will show that whenever the correlation between the catalyst and the external systems is of a quantum-mechanical nature (which we formally specify later), even \textit{arbitrarily small} correlations forbid catalysis, in the sense that the joint state cannot be left unperturbed (Theorem \ref{thm:nocat}).

In doing so, we fully characterize multi-partite states that cannot be used as a catalyst when only local access is allowed. We show that this characterization coincides with the description of quantum states that have no local classical observable (Theorem \ref{thm:main}). Here, classical observables mean those that can be measured without perturbing the global quantum state, in other words, they have non-invasive measurability of deterministically distinguishable states, i.e., obey macro-realism \cite{wilde2012addressing,leggett1985quantum,clemente2016no,pan2020interference}. It turns out that characterizing catalyst with local access is equivalent to characterizing the property known as (tomographical) sensitivity, which was an open problem in the previous work \cite{lie2023faithfulness}.

The observation that quantum correlation that a catalyst has  with other systems  only hinders catalysis leads to the conclusion (Theorem \ref{thm:clcat}) that only classical correlation in correlated catalyst allows for meaningful catalysis. It yields a rather surprising consequence that utilizing a catalyst correlated with an external system is functionally equivalent to using an ensemble of uncorrelated catalysts. In other words, considering correlated catalyst does not introduce new types of non-trivial catalytic transformations, but it only induces probabilistic mixtures of conventional catalytic transformations.

\textit{Background.}--- Recall that a bipartite state $\rho_{AB}$ is said to be classical-quantum (C-Q) when there exists an orthonormal basis $\{\ket{i}\}$ of $A$ such that the following expression is possible:
\begin{equation}
    \rho_{AB}=\sum_i p_i \dyad{i}_A\otimes\rho_B^{(i)},
\end{equation}
with some probability distribution $(p_i)$ and a set of states $\{\rho_B^{(i)}\}$ on $B$. It is equivalent to the existence of a rank-1 projective measurement $\{\dyad{i}\}$ on $A$ that does not disturb the global state $\rho_{AB}$ after the measurement, i.e.,
\begin{equation}
    \rho_{AB}=\sum_{i} (\dyad{i}_A\otimes\mds{1}_B)\rho_{AB}(\dyad{i}_A\otimes\mds{1}_B).
\end{equation}
One can generalize this definition where the projective measurement need not be rank-1 anymore, leading to the following definition of partial classicality.

\begin{definition}\label{def:PCQ}
A bipartite state $\rho_{AB}$ is said to be partially classical-quantum (PC-Q) when there exists a projective measurement $\{\Pi_k\}_{k=1}^n$ with $n>1$ on $A$ that preserves $\rho_{AB}$, i.e.,
\begin{equation}
    \rho_{AB}=\sum_{k=1}^n (\Pi_k\otimes \mds{1}_B)\rho_{AB}(\Pi_k\otimes\mds{1}_B).
\end{equation}
\end{definition}

We will sometimes say that a single system is PC when it is implicitly assumed to be correlated with another system and they are in a PC-Q state. When a system is not PC, then we will say that it is \textit{totally quantum} (TQ) \cite{lie2022delocalized}, so a non PC-Q state is a TQ-Q state.
Note that the correlations in a TQ-Q state can be in general significantly weaker compared to entanglement. 
For example, the following evidently separable state is a TQ-Q state:

\begin{equation} \label{eqn:extq}
    \rho_{AB}=\frac{1}{2}\qty(\sum_{i=0}^{d-1} \lambda_i \dyad{i}_A)\otimes\dyad{0}_B+\frac{1}{2}\dyad{+}_A\otimes\dyad{1}_B,
\end{equation}
where $\sum_{i=0}^{d-1} \lambda_i \dyad{i}_A$ is a nondegenerate quantum state on $A$ and $\ket{+}_A=d^{-1/2}\sum_{i=0}^{d-1}\ket{i}_A$ is a maximally coherent state on $A$. 
This definition of classicality can be further generalized to input or output systems of quantum channels or completely positive (CP) maps: we say that the output system of a quantum channel $\mcal{N}$ is partially classical (PC) when there exists a projective measurement $\mcal{P}:= \sum_k \Pi_k (\cdot )\Pi_k$ that fixes $\mcal{N}$, i.e., $\mcal{P}\circ\mcal{N}=\mcal{N}.$ Similarly we say that the input system of $\mcal{N}$ is PC when a projective measurement $\mcal{Q}$ exists such that $\mcal{N}\circ\mcal{Q}=\mcal{N}$. Likewise, we say that the input or output system is TQ when it is non-PC.

\textit{Totally quantumness and sensitivity.}--- One could question the generality of the notion of totally quantumness, since allowing weaker measurements such as positive operator valued measures (POVM) instead of projective measurements in Definition \ref{def:PCQ} may give rise to a qualitatively different characterization. Perhaps the most natural definition of totally quantumness could be as follows: $A$ of $\rho_{AB}$ is said to be \textit{totally quantum*} (TQ*), or $\rho_{AB}$ is said to be a TQ*-Q state, when any non-trivial quantum measurement on $A$ necessarily perturbs $\rho_{AB}$.
However, by noting that every quantum channel with Kraus operators $\{K_i\}$ can be considered an implementation of the POVM $\{K_i^\dag K_i\}$, we observe that totally quantumness* defined above is equivalent to the concept of (tomographical) sensitivity introduced in Ref.~\cite{lie2023faithfulness}, which characterizes a state's ability to detect the action of \emph{any} non-trivial local channel. 
\begin{definition}
    A bipartite state $\rho_{AB}$ is sensitive on $A$ to a set of quantum operations $\mf{Q}$ with $\id\in\mf{Q}$ when for every $\mcal{S}\in\mf{Q}$
    \begin{equation}
        (\mcal{S}_A\otimes\id_B)(\rho_{AB})=\rho_{AB} \implies \mcal{S}=\id_A.
    \end{equation}
\end{definition}
When a quantum state is sensitive to the set of all quantum channels, we simply say that it is sensitive, or equivalently TQ*-Q. Similarly, through the Choi-\jami isomorphism, we say that a linear map $\mcal{N}$ is sensitive to $\mf{Q}$ when for every $\mcal{S}\in\mf{Q}$
    \begin{equation}\label{eq:channel_sensitivity}
        \mcal{S}\circ\mcal{N}=\mcal{N} \implies \mcal{S}=\id.
    \end{equation}
Our first main result shows that actually the more general definition of totally quantumness is equivalent to the weaker one. This result solves the open problem questioned in Ref. \cite{lie2022delocalized}. 

\begin{theorem} [TQ*=TQ]\label{thm:main}
A quantum channel is sensitive if and only if its output system is TQ. Similarly, a bipartite state $\rho_{AB}$ is sensitive on $A$ if and only if it is a TQ-Q state.
\end{theorem}

Theorem \ref{thm:main} can be shown using the structure result of fixed points of quantum channels \cite{wolf2012quantum}, which in turn follows from the Artin-Wedderburn theorem \cite{artin1927theorie,wedderburn1908hypercomplex}. Self-contained elementary proofs of these results are given in the Supplemental Materials. Theorem \ref{thm:main} says that there is no intermediate level of classicality when it comes to non-invasive measurability; in other words, sensitivity to projective measurements automatically implies sensitivity to general quantum channels. This implies that no classical value can be read from a non-PC system without perturbing it globally, even through weak measurements \cite{brun2002simple,gudder2005non,winter1999coding}.

Theorem \ref{thm:main} yields an interesting property of totally quantumness that it is \textit{contagious}; if any system prepared in a pure state unitarily interacts with a totally quantum system, then it also becomes totally quantum. 
\begin{prop} \label{prop:cont}
    Let $\rho_{AB}$ be a TQ-Q state. For any isometry $V:A\to AK$ such that the marginal state $\tau_E$ of $\tau_{KAB}=(V\otimes \mds{1}_B)\rho_{AB}(V^\dag\otimes \mds{1}_B)$ is full-rank, $\tau_{KAB}$ is a TQ-Q state with respect to the bipartition $K|AB$. 
\end{prop}
An alternative interpretation of Prop. \ref{prop:cont} is that any subsystem of a totally quantum system is also totally quantum. This result is analogous to that of Ref. \cite{streltsov2011linking}, where a local projective measurement on a Q-Q state inevitably forms entanglement with a measurement device. Prop. \ref{prop:cont} shows that this holds similarly even when one considers the more general case of POVMs, and TQ-Q states (proof in Supplemental Material).

 Proposition \ref{prop:cont} shows that it is impossible to circumvent Theorem \ref{thm:main} by unitarily extracting a macro-realistic part $K$ from system $A$ invasively. It means that there could be a non-trivial quantum channel action $\mcal{R}$ on $A$ as a back-action of the invasive measurement, and one can interpret $V$ as the Stinespring dilation of the quantum channel $\mcal{R}$. It provides additional motivation for the nomenclature \textit{totally quantum} system for non-PC systems, as it has no classical property even in the weakest sense, i.e. when classicality means macro-realism and non-invasive measurability.

Theorem \ref{thm:main} offers another intuitive explanation of why quantum key distribution (QKD) is secure. A typical example of TQ-Q state can be found in the BB84 protocol \cite{bennett2014quantum}. When Alice wants to send her random bit (say) 0 to Bob through a quantum channel that could be eavesdropped, she encodes that bit in either of two random bases and record it in her memory $M$,
\begin{equation}
    \rho_{AM}=\frac{1}{2} \dyad{0}_A\otimes\dyad{0}_M + \frac{1}{2} \dyad{+}_A\otimes\dyad{1}_M.
\end{equation}
Since $\rho_{AM}$ is a special case of (\ref{eqn:extq}), by Theorem \ref{thm:main}, any eavesdropper interacting with the qubit $A$ in a non-trivial fashion must alter the global state $
\rho_{AB}$, which results in detectable statistical difference in the later steps of the protocol.

\textit{Local catalysis of bipartite state.}--- Theorem \ref{thm:main} has a significant consequence about catalysis utilizing correlated states. In resource theories, conventionally catalytic transformations mean processes described as
\begin{equation} \label{eqn:cata}
    \rho_S \to \sigma_S:=\Tr_C[\Lambda(\rho_S\otimes\tau_C)],
\end{equation}
with the catalytic constraint requiring that the catalyst remains in its original state in the process:
\begin{equation}
    \Tr_S[\Lambda(\rho_S\otimes\tau_C)] = \tau_C,
\end{equation}
where $C$ is called the catalyst and $\Lambda$ is a free operation on $SC$. 

However, we do not limit ourselves to state transitions between two fixed quantum states. In this work, a catalytic transformation means a general quantum channel $\Phi(\rho)$ given as $\Phi(\rho):=\Tr_C[\Lambda(\rho_S\otimes\tau_C)]$ regardless of whether the initial state $\rho_S$ is fixed or not.
Note that sometimes it is required that the final state of joint system $SC$ has arbitrarily weak correlation, i.e. $\norm{\Lambda(\rho_S\otimes\tau_C)-\sigma_S\otimes\tau_C}_1<\epsilon.$ However, we do not make such an assumption here for generality. 

 Now, using a correlated system $CE$ in state $\tau_{CE}$ as a catalyst when only access to $C$ is given means the transformation of the form in (\ref{eqn:cata}) (where $\tau_S$ is interpreted as  $\Tr_E\tau_{CE}$) with the modified constraint
\begin{equation} \label{eqn:corrcat}
    \Tr_S[\Lambda_{SC}\otimes\id_E(\rho_S\otimes \tau_{CE})]=\tau_{CE}.
\end{equation}

 A typical example of correlated catalyst $\tau_{CE}$ is the product of a previous catalysis, i.e. $\tau_{CE}= \Lambda(\rho_E\otimes \tau_C)$, here $\Lambda$ is same with that in (\ref{eqn:cata}) but acts on $CE$ (See FIG. \ref{fig:TQcatalysis}.) In other words, catalyst $\tau_C$ is `borrowed' by $E$ for catalysis $\Lambda$, formed correlation with $E$, and returned to be borrowed by $S$ again for another round of catalysis. As discussed in Introduction, the bipartite state $\tau_{CE}$ could be used as a resource whenever two systems $C$ and $E$ are combined again, and any change of $\tau_{CE}$ by $S$ can alter its resourceful nature. Our second main result then shows that the catalysis constraint Eq.~(\ref{eqn:corrcat}) severely limits the usability of correlated catalyst.

\begin{theorem} \label{thm:nocat}
    A TQ-Q state $\tau_{CE}$ cannot be used to catalytically implement a non-trivial transformation when only access to $C$ is given.
\end{theorem}
\begin{proof}
    We focus on the fact that once the initial state $\rho_S$ and the interaction channel $\Lambda$ on $SC$ is fixed as in (\ref{eqn:cata}), then the following channel on $C$ is induced.
\begin{equation}
    \Gamma(\eta_C):=\Tr_S[\Lambda(\rho_S\otimes\eta_C)].
\end{equation}
It follows that a catalyst $\tau_{CE}$ must be a fixed point of $\Gamma_C\otimes\id_E$. However, by Theorem \ref{thm:main}, the only channel on $C$ that can fix a TQ-Q $\tau_{CE}$ is the identity channel $\id_C$. It follows that the channel $\Lambda$ on $SC$ must be factorized into $\Lambda_{SC}=\Xi_S\otimes \id_C$. It implies that there is no interaction between $S$ and $C$, therefore the catalysis is trivial.
\end{proof}

 We remark that our proofs did not assume that $\Lambda$ is a free operation, hence the results are applicable to a framework of catalysis much more general than the conventional one where the interaction between system and catalyst should be a free operation. If we assume that $\Lambda_{SC}$ is a free operation, $\Xi_S$ must be free too, if feeding into $\Lambda_{SC}$ a free state on $C$ and discarding it are all free operations. Therefore, the whole transformation is of the form 
$\rho_S \to \Xi_S(\rho_S)$, which is simply a transformation through a free operation.
 
 In principle, catalysts can be also used for non-free operations to reduce the cost (or boost the rate) of transformation, but Theorem \ref{thm:nocat} tells us that not even such generalized catalysis is possible with totally quantum catalysts.

\textit{Mutual information catalysis.}--- If preserving the whole state of $\tau_{CE}$ is too severe a constraint, one might want to preserve only one measure of its correlation, the mutual information $I(C:E)_\tau$. In other words, one might want to catalytically implement the transformation $\rho_S \to \sigma_S := \Tr_{CE}[\Lambda_{SC}\otimes\id_E(\rho_S\otimes\tau_{CE})]$ with the constraint that $I(C:E)_\tau=I(C:E)_\eta,$ where
\begin{equation}
    \eta_{CE}:=\Tr_S[\Lambda_{SC}\otimes\id_E(\rho_S\otimes\tau_{CE})].
\end{equation}
If the above holds, then we say that the transformation $\rho\to\sigma$ is MI(mutual information)-catalytically implemented.

 For this purpose, we first prove the following lemma: if one is required to preserve the mutual information of a TQ-Q state $\rho_{AB}$, then the only actions one can locally apply on the system $A$ are unitary operations. 

\begin{lemma} \label{lem:mip}
    If $\rho_{AB}$ is a TQ-Q state, then for any quantum channel $\mcal{N}$ on $A$ with $\sigma_{AB}:=(\mcal{N}_A\otimes \id_B)(\rho_{AB})$ satisfying $I(A:B)_\rho=I(A:B)_\sigma$ must be a unitary operation.
\end{lemma}
\begin{proof}
    By the data processing inequality, we have $I(A:B)_\rho \geq I(A:B)_\sigma$. By the saturation condition of the data processing inequality, there exists a recovery channel $\mcal{R_N}$ acting on $A$ such that \cite{wilde2013quantum}
\begin{equation}
    ((\mcal{R_N}\circ\mcal{N})_A\otimes\id_B)(\rho_{AB})=\rho_{AB}.
\end{equation}
Since $\rho_{AB}$ is a TQ-Q state, it implies that $\mcal{R_N}\circ\mcal{N}=\id_A$. As the dimensions of input and output systems of $\mcal{N}_A$ are same, it follows that $\mcal{N}_A$ is a unitary operation \cite{nayak2007invertible}.
\end{proof}

By using a similar proof to that of Theorem \ref{thm:nocat}, but substituting the usage of Theorem \ref{thm:main} with Lemma \ref{lem:mip}, we can show that this type of catalysis grants us no additional power either.

\begin{corollary} \label{thm:MIcat}
    A TQ-Q state $\tau_{CE}$ cannot be used to MI-catalytically implement a non-trivial transformation when only access to $C$ is given.
\end{corollary}

This technique provides an answer to the following question: What if different parties try to utilize a multipartite state as a catalyst at the same time? One might wonder if it is possible for two local actions at different sites can cancel each other to enable the recovery of the mutual information. The following result shows that it is nevertheless impossible.
In other words, Corollary \ref{thm:MIcat} explicitly shows that indeed quantum correlation would be a hidden resource; whenever a catalyst is quantumly correlated with an environment, no catalysis is possible without destroying such quantum correlations, as quantified by the mutual information. 

\begin{prop}
For any TQ-Q state $\rho_{AB}$ and two channels $\mcal{N}_A$ and $\mcal{M}_B$, if $(\mcal{N}_A\otimes\mcal{M}_B)(\rho_{AB})=\rho_{AB},$ then $\mcal{N}_A$ is a unitary operation.
\end{prop}

\begin{proof}
Let $\tau_{AB}:=(\mcal{N}_A\otimes\id_B)(\rho_{AB})$ and $\sigma_{AB}:=(\mcal{N}_A\otimes\mcal{M}_B)(\rho_{AB})$. By the data processing inequality, we have $I(A:B)_\rho\geq I(A:B)_\tau \geq I(A:B)_\sigma$. However, as $\sigma_{AB}=\rho_{AB}$, we have $I(A:B)_\rho = I(A:B)_\tau.$ By Lemma \ref{lem:mip}, it follows that $\mcal{N}$ is a unitary operation.
\end{proof}

\textit{General correlated catalysts.}--- One may get the impression that the results above only have implications for a restricted class of bipartite states that are TQ-Q. However, any non TQ-Q state $\tau_{CE}$ is a PC-Q state. In particular, as a consequence of the Koashi-Imoto theorem~\cite{koashi2002operations}, any bipartite state $\tau_{CE}$ can be decomposed into the following form, 
\begin{equation}
    \tau_{CE}=\sum_i p_i \tau_{C_i^L}\otimes\tau_{C_i^RE},
\end{equation}
where $C=\bigoplus_i C_i$ is a direct sum of subspaces $C_i:=C_i^L\otimes C_i^R$ and each $\tau_{C_i^RE}$ is either a TQ-Q state, or $|C_i^R|=1$ (in which case $\tau_{C_i^RE}$ is uncorrelated, but we include it for completeness). See Supplemental Materials for a self-contained and elementary proof of the Koashi-Imoto theorem based on that of Ref. \cite{hayden2004structure}.

This observation tells us how restricted the usage of a general correlated catalyst with local access is. The subspaces $C_i$ of the decomposition above can be interpreted as the `classical degrees of freedom' for $C$ that can be read out without disturbing $\tau_{CE}$, and any quantum channel on $C$ preserving $\tau_{CE}$ also preserve these sectors. It naturally leads us to the following conclusion. (See Supplemental Materials for a detailed discussion.)

\begin{theorem} \label{thm:clcat}
    If a transformation can be catalytically achieved by using the catalyst $\tau_{CE}=\sum_i p_i \tau_{C_i^L}\otimes\tau_{C_i^RE}$ with access to $C$, such that $\tau_{CE}$ is preserved, then the same transformation can be achieved by an ensemble $\qty{p_i,\tau_{C_i^L}}$ of local catalysts.
\end{theorem}

Here, by the ensemble $\qty{p_i,\tau_{C_i^L}}$, we mean the probabilistic mixture of quantum states with classical handle of index $i$, distinguished from the mixed state $\sum_i p_i \tau_{C_i^L}$. Intuitively, using $\tau_{CE}$ on $C$ goes as follows: First, one measures which subspace $C_i$ it is supported on without disturbing $\tau_{CE}$. Since no action can act on $C_i^R$ without disturbing $\tau_{C_i^RE}$, only $\tau_{C_i^L}$ can be utilized as a catalyst. Naturally, it is equivalent to having a local catalyst $\tau_{C_i^L}$ with probability $p_i$. In summary, quantum correlations in correlated catalyst only hinder catalysis, and only the classical part of correlation allows for catalysis because the property of non-invasive measurability, which is essential for recovering the catalyst state.

\textit{Conclusion.} --- 
We inspect the extent to which external correlations that a catalyst might have with its environment (inaccessible to agent) would be affected, when the catalyst is used to facilitate a process. We find that correlations of a quantum-mechanical nature, i.e. contained in TQ-Q states, necessarily degrade in the process of utilizing the catalyst to perform non-trivial transformations. This cautions against potential embezzlement via the consumption of correlations as resources unaccounted for.
Alternatively, from a more constructive viewpoint, our work also shows that there is no advantage of a catalyst bank in creating quantumly correlated states and loaning parts of the states out for catalytic purposes -- they might as well prepare classical ensembles of various independent ancillas.

We emphasize again that the `classicality' here means the non-invasive measurability which means that one can measure a system without altering the state. Hence, our results are not in conflict with previous results on quantum catalysis, where `classicality' indicates other properties, e.g., non-entangled~\cite{horodecki2009quantum}, non-coherent~\cite{streltsov2017colloquium}, non-imaginary~\cite{wu2021resource}, etc. In other words, even when quantum properties of catalyst are utilized, the \textit{state} should be classically known to its user.

\begin{acknowledgements}
\textit{Acknowledgments.} SHL thanks Chae-Yeun Park for helpful discussions on the mathematical proofs. This work was supported by the start-up grant of the Nanyang Assistant Professorship of Nanyang Technological University, Singapore.
\end{acknowledgements}


\appendix
\section{SUPPLEMENTAL MATERIALS}
\section{Notations}
Throughout this Supplemental Materials we will use the following notations. First, every Hilbert space associated with a quantum system is assumed to be finite-dimensional and $|\mcal{H}|$ denotes the dimension of a vector space $\mcal{H}$. The operator space over a Hilbert space $\mcal{H}$ is denoted by $B(\mcal{H})$. We slightly abuse the notation and denote the identity map on $B(\mcal{H})$ by $\id_\mcal{H}$. Also, we will sometimes say that a linear map $\Phi$ defined on $B(\mcal{H})$ is a linear map on $\mcal{H}$, in the sense we identify the operator space associated with a physical system with the system itself. The identity operator in $B(\mcal{H})$, on the other hand, is denoted by $\mds{1}_{\mcal{H}}$. A linear map $\Phi$ on $B(\mcal{H})$ is called completely positive (CP) when $\Phi\otimes\id_\mcal{K}$ is positive for any Hilbert space $\mcal{K}$. A linear map $\Phi$ on $B(\mcal{H})$ is unital if $\Phi(\mds{1}_\mcal{H})=\mds{1}_{\mcal{H}}$.

For any subspace $\mcal{K}$ of a Hilbert space $\mcal{H}$ and a linear map $\Phi$ on $B(\mcal{H})$, we define its limitation $\Phi|_{\mcal{K}}$ as $\Phi$ whose domain is limited to $B(\mcal{K})$ without limiting its image. A `quantum channel on $\mcal{K}$' means a quantum channeldefined on $B(\mcal{K})$ whose image is also in $B(\mcal{K})$.  We will also identify matrices with operators; the term `matrix' will be used to emphasize its algebraic properties. Especially, a `full matrix algebra' is a full operator set over a finite-dimensional Hilbert space, emphasizing that operation composition, i.e. matrix multiplication is well-defined. For any $M\in B(\mcal{H})$, the linear map $\Ad_M$ is given as $\Ad_M(\rho):=M\rho M^\dag$. For any $Q\in B(\mcal{H})$ such that $Q\geq 0$, $\supp{Q}$ is used to denote the support of $Q$, the sum of eigenspaces corresponding to strictly positive eigenvalues of $Q$.

\section{Structure theorem for fixed points of quantum channel}
This section contains the proof of our first main result, i.e. Theorem \ref{thm:main} in the main text. We first provide the proof of a technical result, known in literature \cite{wolf2012quantum} as the structure theorem, and use it subsequently to prove Theorem \ref{thm:main}.

\subsection{Block-diagonal structure}
Of central importance for this section is the structure theorem for the fixed point set of a quantum channel, which we state as S.Theorem \ref{thm:QChstr}.
Many proofs of this theorem relies on the Artin-Wedderburn theorem, where the proof often requires mathematically advanced tools such as ring theory \cite{nicholson1993short, brevsar2010elementary}, functional analysis \cite{wolf2012quantum, hayden2004structure}, or lengthy linear algebraic arguments \cite{koashi2002operations}. In this section, we give an elementary and self-contained proof, by focusing on the more concrete case of our interest.

We begin with the following.
    \begin{slemma}\label{lem:preserving_projection}
        For a unital CP map $\Phi$, if $\Phi(\rho)=\rho$ when $\rho$ is a Hermitian operator with the spectral decomposition $\rho=\sum_{i=1}^n \lambda_i \Pi_i$, we have $\Phi(\Pi_i)=\Pi_i$ for all $i$.
    \end{slemma}
    \begin{proof}
        Without loss of generality, we let $\lambda_i > \lambda_{i+1}$ for all $i$. Let $r_i:=\Tr[\Pi_i]$ and $\pi_i:=\Pi_i/r_i$, then we get that $q_i:=\Tr[\pi_n\Phi(\Pi_i)]$ is a probability distribution. With respect to this distribution, 
        \begin{equation}
            \langle \lambda_i \rangle = \sum_i q_i \lambda_i = \Tr[\pi_n \Phi(\rho)] = \Tr[\pi_n \rho]=\lambda_n.
        \end{equation}
        Since the left hand side is a mean of $\qty{\lambda_i}$ and the right hand side is the smallest element of the set being averaged, it follows that the probability distribution  $\qty{q_i}$ must satisfy that $q_n=1$ and $q_i=0$ for any  $1\leq i \leq n-1$. It implies that $\Phi(\Pi_n)=\Pi_n$, however since $\Phi$ is unital, we also have that 
        \begin{equation}
            \Phi\left(\sum_{i=1}^{n-1}\Pi_i\right)=\sum_{i=1}^{n-1}\Pi_i.
        \end{equation}By limiting $\Phi$ onto the support of $\sum_{i=1}^{n-1}\Pi_i$, we can repeat the same argument and conclude that $\Phi(\Pi_{n-1})=\Pi_{n-1}$. Repeating this process $n$ times gives us the desired result.
    \end{proof}
    \begin{slemma} \label{lem:unicomm}
        For a unital CP map $\Phi$ with Kraus operators $\qty{K_i}$, $\Phi$ fixes $\rho$ if and only if $\qty[K_i,\rho]=\qty[K_i^\dag,\rho]=0$ for all $i$.
    \end{slemma} 
    \begin{proof}
        Showing the ``if" part is trivial; $\Phi(\rho)=\sum_i K_i\rho K_i^\dag = \rho \sum_i K_iK_i^\dag =\rho$. For the other direction, first let us assume that $\rho$ is Hermitian with the spectral decomposition $\rho=\sum_j \lambda_j \Pi_j$. If $\Phi(\rho)=\rho$, then by S.Lemma \ref{lem:preserving_projection} we get that $\Phi(\Pi_j)=\Pi_j$ for all $j$. Next, by conjugating $\Pi_j$ on $\Phi(\Pi_k)=\sum_i K_i \Pi_k K_i^\dag=\Pi_k$, we get 
        \begin{equation}
            \sum_i T^{i}_{jk}T^{i\dag}_{jk}=\delta_{jk}\Pi_k
        \end{equation}where $T^i_{jk}:=\Pi_j K_i \Pi_k$. From the positivity of each $T^{i}_{jk}T^{i\dag}_{jk}$, we get that $T^{i}_{jk}=\Pi_j K_i \Pi_k=0$ for all $i$, whenever $j\neq k$. It means that every $K_i$ is block-diagonal with respect to $\qty{\Pi_j}$;
        \begin{equation}
            K_i\Pi_j=K_i(\mds{1}-\sum_{k\neq j}\Pi_k)=K_i-\sum_{k\neq j} \Pi_k K_i \Pi_k=\Pi_j K_i.
        \end{equation}
        Hence, $\qty[K_i, \Pi_j]=0$ for all $i$ and $j$. Finally, this also then implies that $\qty[K_i,\rho]=0$. By using the fact that a general matrix can be decomposed into Hermitian and anti-Hermitian parts, and a similar analysis for the anti-Hermitian part, we get the desired result.
    \end{proof}
   
    As a side note to complement S.Lemma \ref{lem:unicomm}, we remark that for any set of matrices $\qty{K_i}$, the centralizer of $\qty{K_i,K_i^\dag}$ is the set of fixed points of a unital CP map. This can be shown by downscaling $K_i \to c K_i$ by some constant $c$ such that $\sum_i K_iK_i^\dag \leq \mds{1}$ holds, and using $\qty{K_i}\cup\qty{\sqrt{\mds{1}-\sum_i K_iK_i^\dag}}$ as Kraus operators to construct a unital CP map.\\

    Our main point of interest for this section involves the set of fix points for a unital map $\Phi$, for which we denote as $F_\Phi$. 
    By S.Lemma \ref{lem:unicomm}, we know that if $\rho_1,\rho_2\in F_\Phi$, then their product $\rho_1\rho_2 \in F_\Phi$ also. 
    We remark that because of Hermitian-preserving property of $\Phi$, if $X\in F_\Phi$ then $X^\dag \in F_\Phi$. 
    Furthermore, let us also define $\min F_\Phi$ as the set of \emph{minimal, non-zero central projectors} in $F_\Phi.$ Here, a minimal projector means that it is not a sum of two nonzero projectors, and being central means that it commutes with every element in $F_\Phi$. 
    
    We start with a quick observation on the relation between minimal projectors. Note that for any minimal projector $S$, $SMS=cS$ for some complex number $c$ (we assume that $M$ is Hermitian without loss of generality). Otherwise $SMS$ has a nontrivial spectral decomposition, and a projector onto an eigenspace of $SMS$ will be smaller than $S$, which violates that $S$ is minimal.
    \begin{slemma}\label{lem:minimal_proj_orthogonal}
        If two minimal projectors $S$ and $T$ commute, either $S=T$ or $ST=0$.
    \end{slemma}
\begin{proof}
       This follows from the fact that if $ST\neq 0$ and $S\neq T$, then $ST$ is a projector smaller than $S$, which contradicts that $S$ is minimal.
\end{proof}    
    The next theorem, known as a variant of the Artin-Wedderburn theorem, states that for unital CP maps $\Phi$, the structure of $F_\Phi$ always admits a particular decomposition, i.e. a direct sum according to projections unto $\min F_\Phi$. This particular approach focusing on finding matrix basis elements is inspired by Ref. \cite{brevsar2010elementary}.
    
    \begin{stheorem}[Artin-Wedderburn \cite{artin1927theorie}] For any unital CP map $\Phi$,
    \begin{equation} \label{eqn:cpudec}
        F_\Phi=\bigoplus_{P\in\min F_\Phi} F_\Phi P.
    \end{equation}
        Moreover, there exists a tensor product structure for each $\supp{P}=\mcal{H}_P\otimes\mcal{L}_P$ and  $F_\Phi P$ factorizes into $B(\mcal{H}_P)\otimes \mds{1}_{\mcal{L}_P}$, where $B(\mcal{H}_P)$ is the full matrix algebra on $\mcal{H}_P$.
    \end{stheorem}
    \begin{proof}
        We start by noticing that every two different projectors $P_1,P_2 \in \min F_\Phi$ are orthogonal to each other by S.Lemma \ref{lem:minimal_proj_orthogonal}. Otherwise, $P_1$ cannot be minimal since $P_1=P_1P_2+P_1P_2^{\perp}$ as both $P_1P_2$ and $P_1P_2^\perp$ are central projectors. Next, one can observe that by definition, every central projector in $F_\Phi$ is a sum of minimal central projections. Especially, the identity operator, which is in $F_\Phi$ since $\Phi$, if it is non-minimal, is then also the sum of all projectors in $\min F_\Phi$. Finally, each $F_\Phi P$ is an algebra with $P$ as its unity. 
        
        Now, let us focus on each algebra $F_P:=F_\Phi P$. Although $P$ is a minimal central projector, there could in general be minimal projectors $Q$ in $F_P$ such that $Q<P$, even though they are not central, otherwise P would not be in $\min F_\Phi$. In particular, there exists a decomposition of $P$ into minimal projectors
        \begin{equation}\label{eq:decomp_min_proj}
            P=\sum_i Q_i.
        \end{equation}
 By S.Lemma \ref{lem:minimal_proj_orthogonal}, we also know that all the $Q_i$ in the decomposition of $P$ are mutually orthogonal.   

        Consider the following relation between $Q_i$: $Q_i \sim Q_j$ if there exists $X\in F_P$ such that $Q_i X Q_j \neq 0$. We claim that this relation is an equivalence relation -- the fact that it is reflexive and symmetric is straightforward, while transitivity is also true: if there exists $X$ and $Y$ in $F_P$ such that $Q_i X Q_j \neq 0$ and $Q_j Y Q_k \neq 0$, then
        \begin{equation}
            Z:=Q_i X Q_j Y Q_k \neq 0.
        \end{equation}
        This follows from the fact that 
        \begin{equation}
            Q_j Y Q_k Y^\dag Q_j = c_{jk} Q_j,
        \end{equation}
        with $c_{jk}=\norm{Q_j Y Q_k}^2_2/\Tr[Q_j]$ which is non-zero and similarly, $Q_i X Q_j X^\dag Q_i=c_{ij}Q_i$ with $c_{ij}=\norm{Q_i X Q_j}^2_2/\Tr[Q_i] \neq 0$ so that $ZZ^\dag = c_{ij}c_{jk} Q_i \neq 0$. Therefore the equivalence relation $\sim$ splits $\qty{Q_i}$ into equivalence classes. Moreover, there exists only one equivalence class; Suppose we have two distinct equivalent classes $\mcal{I}$ and $\mcal{J}$ with $Q_\mcal{I}:=\sum \mcal{I}=\sum_{i:Q_i\in\mcal{I}}Q_i$ (similarly for $\mcal{J}$), we then have
        \begin{equation}
            Q_\mcal{I}F_PQ_\mcal{J}=0.
        \end{equation}
        It follows that for any $X\in F_P$, $Q_\mcal{I}X(P-Q_\mcal{I})=0$, hence $Q_\mcal{I}$ becomes central in $F_P$, which contradicts that $P$ is a minimal central projector. As a result, for any two $Q_i$ and $Q_j$, we have $Q_i \sim Q_j$, i.e., there exists $X\in F_P$ such that $Q_i X Q_j \neq 0$.
        
        Now, we let $E_{11}:= Q_1$ and $E_{1j}:=Q_1X_jQ_j$ for some $X_j$, which is guaranteed to exist, such that $\norm{Q_{1j}}_2=\Tr[Q_1 X_j Q_j X_j^\dag]=\Tr[Q_1]$. By letting $X_1:=Q_1$, we can interpret $E_{11}$ as a special case of $E_{1i}$. Then we define $E_{i1}:=E_{1i}^\dag= Q_i X_j^\dag Q_1$ and $E_{ij}:=E_{i1}E_{1j}=Q_i X_{ij} Q_j$ where $X_{ij}:=X_i^\dag Q_1 X_j$ for all $i,j>1$. Because of the property 
        \begin{equation}
            Q_iXQ_i=(\Tr[Q_iX]/\Tr[Q_i])Q_i,
        \end{equation}
        for all $i$ and $X\in F_P$, we have $E_{1i}E_{j1}=Q_1 X_i Q_iQ_j X_j^\dag Q_1=\delta_{ij} \qty(\Tr[Q_1 X_i Q_i X_i^\dag]/\Tr[Q_1]) Q_1=\delta_{ij}Q_1$. Therefore, 
        \begin{equation}
            E_{ij}E_{kl}=E_{i1}(E_{1j}E_{k1})E_{1l}=\delta_{jk}E_{il},
        \end{equation}
        and that $\Tr[E_{ij}^\dag E_{kl}]=\delta_{ik}\delta{jl}$ so that $\norm{E_{ij}}_2^2=\Tr[Q_1]$ for all $i$ and $j$. Especially, $E_{ii}^2=E_{ii}$. Because $E_{ii}=Q_i(X_i^\dag Q_1 X_i) Q_i=r Q_i$ for some positive number $r$, we have $r=1$ so that $E_{ii}=Q_i$. It follows that $\Tr[Q_i]=\Tr[E_{i1}E_{1i}]=\Tr[E_{1i}E_{i1}]=\Tr[Q_1]$ for all $i$, i.e., every minimal projector $Q_i$ has the same rank.

        Next, note that for any $Z\in F_P$, we have $Q_iZQ_j E_{ji}=Q_i(ZQ_j X_{ji})Q_i=c Q_i$ for some complex number $c$ and by taking the trace of the both hands we get $c=\qty(\Tr[E_{ji} Z]/\Tr[Q_1])$ because $Q_jE_{ji}Q_i=E_{ji}$ and $\Tr[Q_i]=\Tr[Q_1]$. Therefore,
        \begin{equation}
            Q_i Z Q_j= \qty(\Tr[E_{ij}^\dag Z]/\Tr[Q_1])E_{ij},
        \end{equation}
        and it follows that for any $Z\in F_P$, $Z=PZP=\sum_{i,j} Q_i Z Q_j= \sum_{i,j} \qty(\Tr[E_{ij}^\dag Z]/\Tr[Q_1]) E_{ij}$, so $\qty{E_{ij}}$ is an orthonormal basis of $F_P$.

        Now, we define a linear map $\Psi$ given as for the basis elements $\qty{E_{ij}}$
        \begin{equation}
            \Psi(E_{ij}):=\dyad{i}{j}_{\mcal{H}_P}\otimes \mds{1}_{\mcal{L}_P},
        \end{equation}
        with some Hilbert spaces $\mcal{H}_P$ and $\mcal{L}_P$ such that $|\mcal{L}_P|=\Tr[Q_1]$. We can see that it is an isomorphism from the fact that $\Psi(E_{ij})\Psi(E_{kl})=\delta_{jk} \Psi(E_{il})$ and $\Tr[\Psi(E_{kl})^\dag \Psi(E_{ij})]=\delta_{ik}\delta_{jl} \Tr[Q_1]=\Tr[E_{kl}^\dag E_{ij}]$. Therefore, one can see that $F_P=F_\Phi P$ is isomorphic to $B(\mcal{H}_P)\otimes\mds{1}_{\mcal{L}_P}$.
    \end{proof}
    For any $\rho_{AB}\geq0$, if $\Tr_B (\rho_{AB})= c\dyad{\psi}_A$ for some pure state $\ket{\psi}_A$ on $A$, then we necessarily have $\rho_{AB}=\dyad{\psi}_A\otimes\rho_B$. Thus, by considering the Choi matrix of each limitation $\Phi|_{\supp{P}}$, we have the following result.

    \begin{scorollary} \label{coro:UCPfac}
        For a unital CP map $\Phi$ with the fixed point set $F_\Phi$ with the decomposition (\ref{eqn:cpudec}), each limitation $\Phi|_{\supp{P}}$ decomposes into
        \begin{equation}
            \Phi|_{\supp{P}}=\id_{\mcal{H}_P}\otimes\Phi_{\mcal{L}_P},
        \end{equation}
        for some unital CP map $\Phi_{\mcal{L}_P}$ on $\mcal{L}_P$ with $\mds{1}_{\mcal{L}_P}$ as the unique fixed point.
    \end{scorollary}
    \begin{proof}
        The set of fixed points of the limitation $\Phi|_\supp{P}$ is $M(\mcal{H}_P)\otimes \mds{1}_{\mcal{L}_P}$. Therefore every Kraus operator of $\Phi|_{\supp{P}}$ commutes with every matrix of the form $A_{\mcal{H}_P}\otimes\mds{1}_{\mcal{L}_P}$, which implies that every Kraus operator is in the form of $\mds{1}_{\mcal{H}_P}\otimes K_i$, hence $\Phi|_{\supp{P}}=\id_{\mcal{H}_P}\otimes\Phi_{\mcal{L}_P}$.
    \end{proof}

   S.Corollary \ref{coro:UCPfac} is a key tool that is necessary for our goal of proving the structure theorem. Additionally, we need a few technical lemmata. We present S.Lemma \ref{lem:LRrank} and \ref{lem:untunq} from Ref. \cite{burgarth2013ergodic} with an alternative proof based on elementary linear algebra. The proofs of S.Lemma \ref{lem:PNprev} and \ref{lem:fpsppf} are taken from Ref. \cite{wolf2012quantum} and presented here for completeness.

    \begin{slemma} [Proposition 6.8, \cite{wolf2012quantum}] \label{lem:PNprev}
        If a quantum channel preserves an operator, then it also preserves the Hermitian and anti-Hermitian part of the operator. Moreover, the channel also preserves the positive and negative parts of the (anti-)Hermitian part.
    \end{slemma}
    \begin{proof}
        Let $\Phi$ be a quantum channel and $\Phi(X)=X$. If $X=H+iA$ where both $H$ and $A$ are Hermitian, then $\Phi(H)-H=i(A-\Phi(A))$. Since the only operator that is both Hermitian and anti-Hermitian is 0, we have $\Phi(H)=H$ and $\Phi(A)=A$. Moreover, if $H=P-N$ where $P,N\geq 0$ and $PN=0$, then by letting $\Pi_P$ be the projector onto $\supp{P}$ and similarly for $N$, we have
        \begin{align}
                &\Tr[P]=\Tr[\Pi_P (P-N)]=\Tr[\Pi_P \Phi(P-N)]\nonumber\\
            \leq& \Tr[\Phi(P)]=\Tr[P].
        \end{align}
    \end{proof}

    \begin{slemma} \label{lem:LRrank}
        If $\lambda$ is an eigenvalue of $A\in B(\mcal{H})$, then the complex conjugate $\lambda^*$ is an eigenvalue of $A^\dag$. Moreover, the geometric multiplicity of $\lambda$ for $A$ is same with that of $\lambda^*$ for $A^\dag$.
    \end{slemma}
    \begin{proof}
        The first part immediately follows from that $\det[A-\lambda\mds{1}]=0$ is equivalent to $\det[A^\dag -\lambda^*\mds{1}]=0$. For the second part, recall that the geometric multiplicity of $\lambda$ for $A$ is equal to $|\mcal{H}|-r(A-\lambda \mds{1})$ where $r(X)$ is the rank of an operator $X$. Because rank is invariant under the adjoint transformation, we have
        \begin{equation}
            |\mcal{H}|-r(A-\lambda \mds{1})=|\mcal{H}|-r(A^\dag-\lambda^* \mds{1}),
        \end{equation}
        and thus we get the wanted result.
    \end{proof}
    
    \begin{slemma} [Proposition 6.10, \cite{wolf2012quantum}] \label{lem:fpsppf}
        For any fixed point $\rho\geq 0$ of a quantum channel $\Phi$, if $Q$ is the projector onto the support of $\rho$, then $\Tr[(\mds{1}-Q)\Phi(Q)]=0$ and
        \begin{equation} \label{eqn:noleak}
            \sigma\leq Q \implies \Phi(\sigma)\leq Q.
        \end{equation}
         Moreover, for projectors $Q$, the condition (\ref{eqn:noleak}) is equivalent to
         \begin{equation}
             \Phi^\dag(Q)\geq Q.
         \end{equation}
         A similar argument can be given for $N$, too.
    \end{slemma}
    \begin{proof}
        For minimal and maximal positive eigenvalues $\lambda_m$ and $\lambda_M$ of $\rho$, we have $\lambda_m Q\leq \rho \leq \lambda_M Q$, hence
        \begin{align*}
            0&\leq\lambda_m\Tr[(\mds{1}-Q)\Phi(Q)]\\&\leq\Tr[(\mds{1}-Q)\Phi(\rho)]\\&=\Tr[(\mds{1}-Q)\rho]\\& \leq \lambda_M\Tr[(\mds{1}-Q)Q]=0.
        \end{align*}
        
        For the second part: \\
        ($\implies$) Let $Q^\perp:=\mds{1}-Q$. Then, 
        \begin{equation}
\Tr[Q^\perp\Phi(Q)]=\Tr[\Phi^\dag(Q^\perp)Q]=0,
        \end{equation}
        hence $Q^\perp\Phi^\dag(Q^\perp)Q^\perp=\Phi^\dag(Q^\perp)$. Using $\Phi(\mds{1})=\mds{1}$, we get
        \begin{equation}
\Phi^\dag(Q)=Q+Q^\perp\Phi^\dag(Q)Q^\perp\geq Q.
        \end{equation}
\noindent     ($\impliedby$) For any $\sigma\leq Q$, we have that 
            \begin{align*}
                \Tr[\sigma]=\Tr[\sigma Q]&=\Tr[\sigma \Phi^\dag(Q)]\\
                &=\Tr[\Phi(\sigma) Q]\\
        &\leq\Tr[\Phi(\sigma)]=\Tr[\sigma]. 
            \end{align*}
            Therefore, $\Tr[\Phi(\sigma) Q]\leq\Tr[\Phi(\sigma)]$, which implies that $\Phi(\sigma)\leq Q$.
    \end{proof}

   \begin{slemma}[Theorem 2, \cite{burgarth2013ergodic}] \label{lem:untunq}
        For any quantum channel $\Phi$ on $A$, if every fixed point of $\Phi^\dag$ is proportional to $\mds{1}_A$, then $\Phi$ also has a unique fixed density matrix $\rho$.
    \end{slemma}
    \begin{proof}
    As a linear map on $B(A)$ (by identifying $A$ with its associated Hilbert space), fixed points of $\Phi$ are equivalent to eigenvectors corresponding to the eigenvalue 1. If every fixed point of $\Phi^\dag$ is proportional to $\mds{1}_A$, then it means that the geometric multiplicity, or the dimension of the eigenspace, of eigenvalue 1 of $\Phi^\dag$ is 1. By S.Lemma \ref{lem:LRrank}, a linear map and its adjoint have the same eigenvalues and the same geometric multiplicities, the geometric multiplicity of 1 as an eigenvalue of $\Phi$ is also 1. It means that there is a unique operator $\rho$ in $B(A)$ such that $\Phi(\rho)=\rho$. This $\rho$ has to be Hermitian because of the Hermitian preserving property of $\Phi$ and moreover, $\rho \geq 0$ because if a Hermitian operator is fixed by a quantum channel, then both of its positive and negative parts should be fixed by the channel by S.Lemma \ref{lem:PNprev}, which makes the geometric multiplicity of 1 larger than 1. After the normalization, it follows that there is only a single quantum state fixed by $\Phi$. 
    \end{proof}

    \begin{stheorem}[Structure theorem for fixed points of quantum channel, \cite{wolf2012quantum}]\label{thm:QChstr}
        For any quantum channel $\Phi$ on $A$, the set $F_\Phi$ of all fixed points of $\Phi$ have the decomposition of the form
        \begin{equation} \label{eqn:qcfpdec}
            F_\Phi=\bigoplus_i B(\mcal{H}_i)\otimes\rho_i.
        \end{equation}
        Here, for any vector space $\mcal{K}$, $\mcal{K}\otimes \rho_i:=\qty{v\otimes \rho_i:v\in\mcal{K}}$.
    \end{stheorem}
    \begin{proof}
        By the previous lemmata, $F_{\Phi^\dag}$ has the decomposition Eq.~\eqref{eqn:cpudec} and for any $P\in\min F_{\Phi^\dag}$, $\Phi^\dag|_{\supp{P}}=\id_{\mcal{H}_P}\otimes \Phi^\dag_{\mcal{L}_P}$. Moreover, since $\Phi^\dag(P)=P$, by S.Lemma \ref{lem:fpsppf}, the image of $\Phi|_{\supp{P}}$ is also contained in $B(\supp{P})$, so that $\qty(\Phi|_{\supp{P}})^\dag=\Phi^\dag|_{\supp{P}}$. It follows that each $\Phi_{\mcal{L}_P}:=\Phi^{\dag\dag}_{\mcal{L}_P}$ is a quantum channel on $\mcal{L}_P$ and has the unique fixed point, say, $\rho_P$, by S.Corollary \ref{coro:UCPfac} and S.Lemma \ref{lem:untunq}. It follows that, for any $X\in B(\mcal{H}_P)$,
        \begin{align*}
             &\Phi(X_{\mcal{H}_P}\otimes \rho_P)=\Phi|_{\mcal{H}_P\otimes\mcal{L}_P}(X_{\mcal{H}_P}\otimes \rho_P)\\
            =&\id_{\mcal{H}_P}(X)\otimes \Phi_{\mcal{L}_P}(\rho_P)=X_{\mcal{H}_P}\otimes \rho_P.
        \end{align*}
        In other words, $B(\mcal{H}_P)\otimes \rho_P\subseteq F_\Phi$. Therefore, the following direct sum
        \begin{equation}
            G_\Phi:=\bigoplus_{P\in \min F_\Phi^\dag} B(\mcal{H}_P)\otimes \rho_P,
        \end{equation}
        is a subspace of $F_\Phi$ as a vector space over the complex number field, because each summand is fixed by $\Phi$. 
        
        Now, we count the dimension of each space. Recall that the dimension of a direct sum of vector spaces is the sum of the dimension of all the individual summand. Because $|B(\mcal{H}_P)\otimes \rho_P|=|B(\mcal{H})||\,\text{span}\qty{\rho_P}|=|B(\mcal{H})|$ as $\rho_P$ is understood as a single-point set and the span of it is 1-dimensional, we have
        \begin{equation}
            |G_\Phi|=\sum_{P\in\min F_{\Phi^\dag}} |B(\mcal{H}_P)|=\sum_{P\in \min F_{\Phi^\dag}} |\mcal{H}_P|^2.
        \end{equation}
         On the other hand, similarly $|B(\mcal{H}_P)\otimes \mds{1}_{\mcal{L}_P}|=|B(\mcal{H}_P)|$, therefore it follows that
        \begin{equation}
            |F_{\Phi^\dag}|=\sum_{P\in\min F_{\Phi^\dag}} |B(\mcal{H}_P)|=\sum_{P\in \min F_{\Phi^\dag}} |\mcal{H}_P|^2.
        \end{equation}
        Finally, note that $|F_\Phi|=|F_{\Phi^\dag}|$ by S.Lemma \ref{lem:LRrank}, because they correspond to the geometric multiplicities of 1 for $\Phi$ and $\Phi^\dag$, respectively. It follows that $G_\Phi=F_\Phi$ because their dimensions are the same.
    \end{proof}

\subsection{Proof of Theorem \ref{thm:main}}

\begin{proof}
The first part of the theorem is straightforward by looking at the contrapositive: if a quantum channel $\mathcal{N}$ has an output system that is not totally quantum, i.e. PC, then by Eq.~\eqref{eq:channel_sensitivity} there exists a non-trivial projective measurement, which is a quantum channel that fixes $\mathcal{N}$, hence $\mathcal{N}$ cannot be sensitive.

We show the other direction. Let the output system of $\mcal{N}$ be TQ and let us say that a quantum channel $\mcal{S}$ fixes $\mcal{N}$, i.e., $\mcal{S}\circ\mcal{N}=\mcal{N}.$ This means that the image of $\mcal{N}$ is a subset of fixed points of $\mcal{S}$. By S.Theorem \ref{thm:QChstr}, the space of all fixed points $F_\mcal{S}$ of $\mcal{S}$ must have the following unique decomposition,
\begin{equation} \label{eqn:directsum}
    F_\mcal{S}=\bigoplus_i \mcal{M}_{d_i}\otimes \sigma_i ,
\end{equation}
with respect to an appropriate basis and for some fixed full-rank quantum states $\sigma_i$ and the full matrix algebras $\mcal{M}_{d_i}$.

If there is more than one term in the direct sum of Eq.~\eqref{eqn:directsum}, we will call the set of projectors onto their supports $\{\Pi_i\}$. Note that the projective measurement described by $\{\Pi_i\}$ fixes all the states in
$ F_\mathcal{S}$, and therefore all the output states of $\mcal{N}$. This contradicts the assumption that the output system of $\mcal{N}$ is TQ. Hence, there should be only one term in the direct sum of Eq.~\eqref{eqn:directsum} so that $F_\mcal{S}$ should have the form of $F_\mcal{S}=\mcal{M}_d\otimes \sigma$ for some full matrix algebra $\mcal{M}_d$ and a fixed quantum state $\sigma$. However, suppose that there are more than one distinct eigenvalue for $\sigma$ so that there are at least two terms in the spectral decomposition of $\sigma=\sum_j \lambda_j P_j$, where $P_j$ is the projector onto the eigenspace corresponding to the eigenvalue $\lambda_j$. Then, we can now identify a projective measurement that fixes $F_\mcal{S}$, i.e.  $\{\mds{1}_d\otimes P_j\}$ where $\mds{1}_d$ is the identity matrix in $\mcal{M}_d$. It also violates the output system of $\mcal{N}$ being non-PC, hence $\sigma$ should be a 1-dimensional state so that $F_\mcal{S}$ is isomorphic to a full matrix algebra, i.e., $\mcal{S}$ fixes \emph{every} operator. This means that $\mcal{S}=\id$ and hence $\mcal{N}$ is sensitive.

The statement for bipartite states follows the proof above by applying the Choi-\jami isomorphism to $\mcal{S}$. 
\end{proof}

\section{Proof of Proposition \ref{prop:cont}}
\begin{proof}
    Let us start by noting that for any isometry $V:A\to AK$, there exists a unitary operator $U$ on $AK$ such that $V=U(\ket{\phi}_K\otimes \mds{1}_A)$ for some state $\ket{\phi}_K$. Additionally, consider a measurement operation $\mcal{P}_K(\cdot)=\sum_i \Pi_i \cdot \Pi_i$ with more than one mutually orthogonal projectors $\{\Pi_i\}$ on $K$ that fixes $\tau_{KAB}$. Consider an induced quantum channel on $A$ defined as
    \begin{equation}
        \mcal{Q}(\cdot):=\Tr_K[\mcal{U}^\dag\circ \mcal{P}_K\circ\mcal{U}(\dyad{\phi}_K\otimes\cdot_A)]
    \end{equation}
    where $\mcal{U}$ is the unitary channel describing the action of $U$ on $AK$. Since $\mcal{P}_E$ fixes $\tau_{KAB}$, $\mcal{Q}_A$ fixes $\rho_{AB}$. Since it is a TQ-Q state, it implies that $\mcal{Q}$ is the identity channel. Since unitary operations cannot form correlation with other systems, (or by the Schr\"{o}dinger-HJW theorem \cite{schrodinger1935discussion,hughston1993complete}) that $\mcal{U}^\dag\circ \mcal{P}_K\circ\mcal{U}(\dyad{\phi}_K\otimes\id_A)=\sigma_K\otimes\id_A$ for some state $\sigma_K$. Applying this channel to $\rho_{AB}$ on $A$, we get
    \begin{equation}
        \mcal{U}_{KA}^\dag\circ\mcal{P}_E\circ\mcal{U}_{KA}(\dyad{\phi}_K\otimes\rho_{AB})=\sigma_{K}\otimes\rho_{AB}.
    \end{equation}
    Note that $\mcal{U}_{KA}(\dyad{\phi}_K\otimes\rho_{AB})=\tau_{KAB}$ and $\mcal{P}_K$ does not alter $\tau_{KAB}$. Hence, the left hand side is $\mcal{U}_{KA}^\dag\circ\mcal{U}_{KA}(\dyad{\phi}_K\otimes\rho_{AB})=\dyad{\phi}_K\otimes\rho_{AB}$. It follows that $\sigma_K=\dyad{\phi}_K$.
    
    Now, a quantum state $\eta$ is fixed by measurement operation $\mcal{P}(\cdot)=\sum_i \Pi_i \cdot \Pi_i$ if and only if $\Pi_i \eta \Pi_i=\eta$ for only one $i$. Moreover, $\mcal{P}_K\otimes \id_A$ is also a measurement operation. Considering the Choi matrix, it follows that there exists a unique $\Pi_i$ such that $(\Pi_i\otimes\mds{1}_A)U(\ket{\phi}_K\otimes\mds{1}_A)=U(\ket{\phi}_K\otimes \mds{1}_A)$ and $(\Pi_j\otimes\mds{1}_A)U(\ket{\phi}_K\otimes\mds{1}_A)=0$ for any other $\Pi_j$. By conjugating the operator above to $\rho_{AB}$ on $A$ and tracing out $B$, we get that $\Pi_j\tau_K\Pi_j=0$, which contradicts $\tau_A$ being full-rank.
\end{proof}

\section{Proof of Theorem \ref{thm:clcat}}

\subsection{The Koashi-Imoto Theorem and Decomposition of PC-Q state}
Here, we prove that arbitrary bipartite state $\tau_{CE}$ can be decomposed into the from
\begin{equation} \label{eqn:PCdec}
    \tau_{CE}=\bigoplus_i p_i \tau_{C_i^L}\otimes\tau_{C_i^RE}
\end{equation}
where $C=\bigoplus_i C_i^L\otimes C_i^R$ and each $\tau_{C_i^RE}$ is a TQ-Q state. Equivalently, through the Choi-\jami isomorphism, for any quantum channel $\mcal{T}:E\to C$, one has the following unique decomposition.
\begin{equation} \label{eqn:qcdec}
    \mcal{T}(\rho)=\bigoplus_i \tau_{C_i^L}\otimes\mcal{T}_{C_i^R}(\rho).
\end{equation}
with respect to the same $C$ above for any input state $\rho$, where $\tau_{C_i^L}$ is a quantum state on $C_i^L$ and $\mcal{T}_{C_i^R}:E\to C_i^R$ is a sensitive, trace non-increasing~\footnote{This is also why the probabilities in Eq.~\eqref{eqn:PCdec} were ommitted in Eq.~\eqref{eqn:qcdec}, since w.l.o.g. they can be absorbed into $\mathcal{T}_{C_i^R}$.} CP map (or a trivial map with $|C_i^R|=1$, which we simply count as a special case of sensitive map for simplicity). This result can be attained by using the Koashi-Imoto theorem \cite{koashi2002operations} as a lemma. Here, we provide a concise statement and proof of the Koashi-Imoto theorem that mainly follows that of Ref. \cite{hayden2004structure} without using the result of Ref. \cite{takesaki2003theory} directly, by using the tools that facilitated the structure theorem (S.Theorem \ref{thm:QChstr}) we developed earlier.

\begin{slemma}[Koashi-Imoto \cite{koashi2002operations, hayden2004structure}]\label{lem:koim}
    For any set of quantum states $\qty{\rho_k}$ on a Hilbert space $\mcal{H}$, there exists a unique decomposition of $\mcal{H}=\bigoplus_i \mcal{H}_{i^L}\otimes\mcal{H}_{i^R}$ that satisfies the following:
    \begin{enumerate}[label=(\roman*),leftmargin=15pt]
        \item Each $\rho_k$ decomposes as
    \begin{equation} \label{eqn:KIdec}
        \rho_k=\bigoplus_i q_{i|k} ~\omega_{i^L} \otimes  \rho_{i^R|k},
    \end{equation}
    where $(q_{i|k})$ is a probability distribution over $i$ and $\rho_{i^R|k}$ is a quantum state on $\mcal{H}_{i^R}$ depending on $k$, while $\omega_{i^L}$ is a quantum state on $\mcal{H}_{i^L}$ independent of $k$.
    \item Any quantum channel $\mcal{C}$ on $\mcal{H}$ that fixes all $\rho_k$ is a quantum channel on each subspace $\mcal{H}_{i^L}\otimes\mcal{H}_{i^R}$ and 
    \begin{equation} \label{eqn:fixcde}
        \mcal{C}|_{\mcal{H}_{i^L}\otimes\mcal{H}_{i^R}}= \mcal{C}_{\mcal{H}_{i^L}}\otimes \id_{\mcal{H}_{i^R}}, \qquad \forall i,
    \end{equation}
 where each $\mcal{C}_{\mcal{H}_{i^L}}$ fixes $\omega_{i^L}$, i.e., $\mcal{C}_{\mcal{H}_{i^L}}(\omega_{i^L})=\omega_{i^L}$.
    \end{enumerate}
\end{slemma}

\begin{proof}
    Let $\mathbf{F}:=\qty{\mcal{F}:\mcal{F}(\rho_k)=\rho_k, \forall k}$ be the set of all quantum channels that fixes each $\qty{\rho_k}$. Then, when $F_{\mcal{C}}$ is the set of all fixed points of a linear map $\mcal{C}$, we let
    \begin{equation}
        F_{0}:=\bigcap_{\mcal{F}\in \mathbf{F}} F_{\mcal{F}^\dag}.
    \end{equation}
    Since every $F_{\mcal{F}^\dag}$ is finite dimensional, $F_{0}$ can be actually expressed as a finite intersection:
    \begin{equation}
        F_0 = F_{\mcal{F}_1^\dag}\cap F_{\mcal{F}_2^\dag}\cap \cdots \cap F_{\mcal{F}_M^\dag}, 
    \end{equation}
    for some $\mcal{F}_1,\cdots,\mcal{F}_M\in \mathbf{F}$. To see this, observe that intersecting one more $F_\mcal{F}$ can never increase the dimension of the intersection, hence becasue of the finite dimensionality, only a finite number of $F_{\mcal{F}_i^\dag}$ nontrivially affect the interaction $\bigcap_{\mcal{F}\in \mathbf{F}} F_{\mcal{F}^\dag}$. Let us consider the quantum channel $\mcal{F}_0$ given as
    \begin{equation}
    \mcal{F}_0:=\frac{1}{M}\sum_{n=1}^M \mcal{F}_n.
    \end{equation}
    The Kraus operators of $\mcal{F}_0^\dag$ are simply the union of scalar multiples of those of $\mcal{F}_i^\dag$. Suppose now, that we have a state $\rho\in F_{\mcal{F}_0^\dag}$. Then, by S.Lemma \ref{lem:unicomm}, $\rho$ commutes with all the Kraus operators of each $\mathcal{F}_i^\dag$. Therefore, again by S.Lemma \ref{lem:unicomm}, $\rho$ is fixed by all of $\mcal{F}_i^\dag$, i.e., $\rho\in F_{\mcal{F}_1^\dag}\cap F_{\mcal{F}_2^\dag}\cap \cdots \cap F_{\mcal{F}_M^\dag}$. Hence $\rho \in F_0$. It shows that $F_{\mcal{F}_0^\dag}\subseteq F_0$, and therefore $F_0 = F_{\mcal{F}_0^\dag}$.
    It follows that there exists the decomposition of the form \eqref{eqn:cpudec}
    \begin{equation} \label{eqn:FFdec}
        F_{\mcal{F}_0^\dag}=\bigoplus_i \mds{1}_{\mcal{H}_{i^L}}\otimes B(\mcal{H}_{i^R}),
    \end{equation}
    with respect to a decomposition $\mcal{H}=\bigoplus_i \mcal{H}_{i^L}\otimes\mcal{H}_{i^R}$. Define $Q_i:=\mds{1}_{\mcal{H}_{i^L}}\otimes\mds{1}_{\mcal{H}_{i^R}}$ to be the projector onto the subspace $\mcal{H}_{i^L}\otimes\mcal{H}_{i^R}$ of $\mcal{H}$. Then, we can observe from the block-diagonal structure (\ref{eqn:FFdec}) that
    \begin{equation}
        Q_i ~ \in  F_{\mcal{F}_0^\dag}=\bigcap_{\mcal{G}\in \mathbf{F}} F_{\mcal{G}^\dag}\subseteq F_{\mcal{F}^\dag}
    \end{equation}
    for any $\mcal{F}\in \mathbf{F}$. It means that $\mcal{F}^\dag(Q_i)=Q_i$ and thus  $\mcal{F}^\dag(Q_i)\geq Q_i$ and $\mcal{F}^\dag(\mds{1}-Q_i)\geq \mds{1}-Q_i$ for any $\mcal{F}\in\mathbf{F}$ and $i$ because $\mcal{F}^\dag(\mds{1})=\mds{1}$. By S.Lemma \ref{lem:fpsppf}, it follows that, for all $i$ and $\mcal{F}\in\mathbf{F}$, $\mcal{F}|_{\mcal{H}_{i^L}\otimes\mcal{H}_{i^R}}$ is a quantum channel on $\mcal{H}_{i^L}\otimes\mcal{H}_{i^R}$ and $\qty(\mcal{F}|_{\mcal{H}_{i^L}\otimes\mcal{H}_{i^R}})^\dag = \mcal{F}^\dag|_{\mcal{H}_{i^L}\otimes\mcal{H}_{i^R}}$. 
    
    Especially for $\mathcal{F}_0$, from (\ref{eqn:FFdec}) it follows that
    \begin{equation}        \mcal{F}_0^\dag|_{\mcal{H}_{i^L}\otimes\mcal{H}_{i^R}}=\mcal{G}_{0\mcal{H}_{i^L}}\otimes \id_{\mcal{H}_{i^R}},
    \end{equation}
    with some unital CP map $\mcal{G}_{0\mcal{H}_{i^L}}$ on $\mcal{H}_{i^L}$ with the one-dimensional fixed point set $\qty{c\mds{1}}$. By S.Lemma \ref{lem:untunq}, 
    \begin{equation}        \mcal{F}_0|_{\mcal{H}_{i^L}\otimes\mcal{H}_{i^R}}=\mcal{F}_{0\mcal{H}_{i^L}}\otimes \id_{\mcal{H}_{i^R}},
    \end{equation}
    with some quantum channel $\mcal{F}_{0\mcal{H}_{i^L}}$ on $\mcal{H}_{i^L}$ with a unique fixed quantum state, say, $\omega_{i^L}$. Therefore,
    \begin{equation}
        F_{\mcal{F}_0}=\bigoplus_i \omega_{i^L}\otimes B(\mcal{H}_{i^R}).
    \end{equation}
    It follows that every $\rho_k$, as a fixed point of $\mcal{F}_0$, indeed has the expression of the form (\ref{eqn:KIdec}).

    Now we prove $(ii)$. For any $\mcal{F}\in\mathbf{F}$, again, $\mcal{F}^\dag|_{\mcal{H}_{i^L}\otimes\mcal{H}_{i^R}}=\mcal{F}_{\mcal{H}_{i^L}}^\dag\otimes\id_{\mcal{H}_{i^R}}$ for some quantum channel $\mcal{F}_{\mcal{H}_{i^L}}$ on $\mcal{H}_{i^L}$, because $F_0=F_{\mcal{F}_0^\dag}\subseteq F_{\mcal{F}^\dag}$, so $\mcal{F}|_{\mcal{H}_{i^L}\otimes\mcal{H}_{i^R}}=\mcal{F}_{\mcal{H}_{i^L}}\otimes\id_{\mcal{H}_{i^R}}$. Since $\qty{\rho_k}\subseteq F_{\mcal{F}}$, it must be that $\omega_{i^L}\otimes\rho_{i^R|k}$ is a fixed point of $\mcal{F}_{\mcal{H}_{i^L}}\otimes \id_{\mcal{H}_{i^R}}$ for all $i$. Thus,  every $\mcal{F}_{\mcal{H}_{i^L}}$ must have $\omega_{i^L}$ as a fixed state. It proves $(ii)$.

\end{proof}

We apply this result on the image of the quantum channel $\mcal{T}$, $\qty{\mcal{T}(\rho)}$ so that the input state $\rho$ functions as the index $k$ in the statement of the Koashi-Imoto theorem above. Hence, according to $(i)$ above, for any $\rho$, $\mcal{T}(\rho)$ decomposes as
\begin{equation} 
    \mcal{T}(\rho)=\bigoplus_i \omega_{i^L}\otimes\mcal{T}_{i^R}(\rho).
\end{equation}
Since $\omega_{i^L}$ is independent of $\rho$, it follows that each $\mcal{T}_{i^R}$ is linear in $\rho$. Moreover, it is immediate that they are CP. Now we claim that each $\mcal{T}_{i^R}$ is sensitive, because, otherwise, there exists a non-identity quantum channel $\mcal{N}_{i^R}$ on $\mcal{H}_{i^R}$ such that $\mcal{N}_{i^R}\circ\mcal{T}_{i^R}=\mcal{T}_{i^R}$ and it contradicts $(ii)$ of Lemma \ref{lem:koim}.

Once we have the decomposition of arbitrary quantum channel (\ref{eqn:qcdec}), by using the Choi-\jami isomorphism again, for each CP map $\mcal{T}_{i^R}$ one can find the corresponding bipartite quantum state $\rho_{C_i^R E}$ and normalization factor $q_i:=\Tr[\mcal{T}_{i^R}(\mds{1}_\mcal{H})]/|\mcal{H}|$ to get (\ref{eqn:PCdec}), which should sum up to 1 due to the trace preserving property of $\mcal{T}$. In this case, for any quantum channel $\mcal{C}$ on $C$ such that $\mcal{C}_C\otimes\id_E(\rho_{CE})=\rho_{CE}$, $\mcal{C}$ should satisfy (\ref{eqn:fixcde}).

\subsection{Proof of Theorem \ref{thm:clcat}}

We follow the logic of the proof of Theorem \ref{thm:main} above. Consider any interaction $\Lambda$ on $SC$ that implements the catalytic transformation of a state $\rho_S$ on $S$ using $\tau_{CE}$ with access to $C$. 
We consider the channel $\mcal{S}$ on $S$, defined as
\begin{equation}
    \mcal{S}(\rho):=\Tr_{CE}[\Lambda_{SC}\otimes\id_E(\rho_S\otimes\tau_{CE})].
\end{equation}
Recall the statement of the theorem presumes the following decomposition of the catalyst with its environment:
\begin{equation}
    \tau_{CE}=\sum_i p_i \tau_{C_i^L}\otimes\tau_{C_i^RE}.
\end{equation}
Let $\mcal{P}=\sum_j \Pi_j \cdot \Pi_j$ be the pinching map on $C$ where each $\Pi_j$ is the projector onto the subspace $C_j^L\otimes C_j^R$. This models the process that acts locally on the system after a projective measurement processes. Note that $\mcal{P}_C\otimes\id_E$ fixes $\tau_{CE}$, i.e. 
\begin{equation}
    \mcal{P}_C\otimes\id_E (\tau_{CE}) = \tau_{CE}.
\end{equation}Thus, one can perform the pinching operation before applying $\Lambda_{SC}$ without changing $\mcal{S}$:
\begin{align}
    &\Tr_{CE}[\Lambda_{SC}\otimes\id_E(\rho_S\otimes\tau_{CE})]\nonumber\\
&=\Tr_{CE}[\Lambda_{SC}\circ(\id_{SE}\otimes\mcal{P}_C)(\rho_S\otimes\tau_{CE})]\\
&=\Tr_C[\Lambda_{SC}\circ(\id_S \otimes \mcal{P}_C)(\rho_S\otimes\tau_C)]\nonumber.
\end{align}
Hence, from the discussion of the previous section, we get the factorization
\begin{equation}
    \Lambda_{SC}(\rho_S\otimes\Pi_j\tau_C\Pi_j)=p_j\Lambda_{SC_j^L}(\rho_S\otimes\tau_{C_j^L})\otimes \tau_{C_j^R}
\end{equation}
with the limitation $\Lambda_{SC_j^L}$ of $\Lambda_{SC}$ onto $SC_j^L$ which is by itself a quantum channel on $SC_j^L$. The resultant transformation of $\rho_S$ is therefore of the form
\begin{equation}\label{eq:mcalS}
    \mcal{S}(\rho)=\sum_j p_j \Tr_C[\Lambda_{SC_j^L}(\rho_S\otimes\tau_{C_j^L})]
\end{equation}Therefore, one can see that $\mcal{S}$ can be implemented with the ensemble $\qty{p_i,\tau_{C_i^L}}$.

Conversely, any catalysis possible with the ensemble $\qty{p_i,\tau_{C_i^L}}$ is also possible with $\tau_{CE}$ through the protocol explained in the main text after Theorem \ref{thm:clcat}.

\bibliography{main}

\begin{thebibliography}{48}%
\makeatletter
\providecommand \@ifxundefined [1]{%
 \@ifx{#1\undefined}
}%
\providecommand \@ifnum [1]{%
 \ifnum #1\expandafter \@firstoftwo
 \else \expandafter \@secondoftwo
 \fi
}%
\providecommand \@ifx [1]{%
 \ifx #1\expandafter \@firstoftwo
 \else \expandafter \@secondoftwo
 \fi
}%
\providecommand \natexlab [1]{#1}%
\providecommand \enquote  [1]{``#1''}%
\providecommand \bibnamefont  [1]{#1}%
\providecommand \bibfnamefont [1]{#1}%
\providecommand \citenamefont [1]{#1}%
\providecommand \href@noop [0]{\@secondoftwo}%
\providecommand \href [0]{\begingroup \@sanitize@url \@href}%
\providecommand \@href[1]{\@@startlink{#1}\@@href}%
\providecommand \@@href[1]{\endgroup#1\@@endlink}%
\providecommand \@sanitize@url [0]{\catcode `\\12\catcode `\$12\catcode
  `\&12\catcode `\#12\catcode `\^12\catcode `\_12\catcode `\%12\relax}%
\providecommand \@@startlink[1]{}%
\providecommand \@@endlink[0]{}%
\providecommand \url  [0]{\begingroup\@sanitize@url \@url }%
\providecommand \@url [1]{\endgroup\@href {#1}{\urlprefix }}%
\providecommand \urlprefix  [0]{URL }%
\providecommand \Eprint [0]{\href }%
\providecommand \doibase [0]{http://dx.doi.org/}%
\providecommand \selectlanguage [0]{\@gobble}%
\providecommand \bibinfo  [0]{\@secondoftwo}%
\providecommand \bibfield  [0]{\@secondoftwo}%
\providecommand \translation [1]{[#1]}%
\providecommand \BibitemOpen [0]{}%
\providecommand \bibitemStop [0]{}%
\providecommand \bibitemNoStop [0]{.\EOS\space}%
\providecommand \EOS [0]{\spacefactor3000\relax}%
\providecommand \BibitemShut  [1]{\csname bibitem#1\endcsname}%
\let\auto@bib@innerbib\@empty
\bibitem [{\citenamefont {Datta}\ \emph {et~al.}(2022)\citenamefont {Datta},
  \citenamefont {Varun~Kondra}, \citenamefont {Miller},\ and\ \citenamefont
  {Streltsov}}]{datta2022catalysis}%
  \BibitemOpen
  \bibfield  {author} {\bibinfo {author} {\bibfnamefont {C.}~\bibnamefont
  {Datta}}, \bibinfo {author} {\bibfnamefont {T.}~\bibnamefont {Varun~Kondra}},
  \bibinfo {author} {\bibfnamefont {M.}~\bibnamefont {Miller}}, \ and\ \bibinfo
  {author} {\bibfnamefont {A.}~\bibnamefont {Streltsov}},\ }\href@noop {}
  {\bibfield  {journal} {\bibinfo  {journal} {arXiv e-prints}\ ,\ \bibinfo
  {pages} {arXiv}} (\bibinfo {year} {2022})}\BibitemShut {NoStop}%
\bibitem [{\citenamefont {Jonathan}\ and\ \citenamefont
  {Plenio}(1999)}]{jonathan1999entanglement}%
  \BibitemOpen
  \bibfield  {author} {\bibinfo {author} {\bibfnamefont {D.}~\bibnamefont
  {Jonathan}}\ and\ \bibinfo {author} {\bibfnamefont {M.~B.}\ \bibnamefont
  {Plenio}},\ }\href@noop {} {\bibfield  {journal} {\bibinfo  {journal}
  {Physical Review Letters}\ }\textbf {\bibinfo {volume} {83}},\ \bibinfo
  {pages} {3566} (\bibinfo {year} {1999})}\BibitemShut {NoStop}%
\bibitem [{\citenamefont {Daftuar}\ and\ \citenamefont
  {Klimesh}(2001)}]{daftuar2001mathematical}%
  \BibitemOpen
  \bibfield  {author} {\bibinfo {author} {\bibfnamefont {S.}~\bibnamefont
  {Daftuar}}\ and\ \bibinfo {author} {\bibfnamefont {M.}~\bibnamefont
  {Klimesh}},\ }\href@noop {} {\bibfield  {journal} {\bibinfo  {journal}
  {Physical Review A}\ }\textbf {\bibinfo {volume} {64}},\ \bibinfo {pages}
  {042314} (\bibinfo {year} {2001})}\BibitemShut {NoStop}%
\bibitem [{\citenamefont {Klimesh}(2007)}]{klimesh2007inequalities}%
  \BibitemOpen
  \bibfield  {author} {\bibinfo {author} {\bibfnamefont {M.}~\bibnamefont
  {Klimesh}},\ }\href@noop {} {\bibfield  {journal} {\bibinfo  {journal} {arXiv
  preprint arXiv:0709.3680}\ } (\bibinfo {year} {2007})}\BibitemShut {NoStop}%
\bibitem [{\citenamefont {Aubrun}\ and\ \citenamefont
  {Nechita}(2008)}]{aubrun2008catalytic}%
  \BibitemOpen
  \bibfield  {author} {\bibinfo {author} {\bibfnamefont {G.}~\bibnamefont
  {Aubrun}}\ and\ \bibinfo {author} {\bibfnamefont {I.}~\bibnamefont
  {Nechita}},\ }\href@noop {} {\bibfield  {journal} {\bibinfo  {journal}
  {Communications in Mathematical Physics}\ }\textbf {\bibinfo {volume}
  {278}},\ \bibinfo {pages} {133} (\bibinfo {year} {2008})}\BibitemShut
  {NoStop}%
\bibitem [{\citenamefont {Campbell}(2011)}]{campbell2011catalysis}%
  \BibitemOpen
  \bibfield  {author} {\bibinfo {author} {\bibfnamefont {E.~T.}\ \bibnamefont
  {Campbell}},\ }\href@noop {} {\bibfield  {journal} {\bibinfo  {journal}
  {Physical Review A}\ }\textbf {\bibinfo {volume} {83}},\ \bibinfo {pages}
  {032317} (\bibinfo {year} {2011})}\BibitemShut {NoStop}%
\bibitem [{\citenamefont {Brandao}\ \emph {et~al.}(2015)\citenamefont
  {Brandao}, \citenamefont {Horodecki}, \citenamefont {Ng}, \citenamefont
  {Oppenheim},\ and\ \citenamefont {Wehner}}]{brandao2015second}%
  \BibitemOpen
  \bibfield  {author} {\bibinfo {author} {\bibfnamefont {F.}~\bibnamefont
  {Brandao}}, \bibinfo {author} {\bibfnamefont {M.}~\bibnamefont {Horodecki}},
  \bibinfo {author} {\bibfnamefont {N.}~\bibnamefont {Ng}}, \bibinfo {author}
  {\bibfnamefont {J.}~\bibnamefont {Oppenheim}}, \ and\ \bibinfo {author}
  {\bibfnamefont {S.}~\bibnamefont {Wehner}},\ }\href@noop {} {\bibfield
  {journal} {\bibinfo  {journal} {Proceedings of the National Academy of
  Sciences}\ }\textbf {\bibinfo {volume} {112}},\ \bibinfo {pages} {3275}
  (\bibinfo {year} {2015})}\BibitemShut {NoStop}%
\bibitem [{\citenamefont {{\AA}berg}(2014)}]{aaberg2014catalytic}%
  \BibitemOpen
  \bibfield  {author} {\bibinfo {author} {\bibfnamefont {J.}~\bibnamefont
  {{\AA}berg}},\ }\href@noop {} {\bibfield  {journal} {\bibinfo  {journal}
  {Physical Review Letters}\ }\textbf {\bibinfo {volume} {113}},\ \bibinfo
  {pages} {150402} (\bibinfo {year} {2014})}\BibitemShut {NoStop}%
\bibitem [{\citenamefont {M{\"u}ller}(2018)}]{muller2018correlating}%
  \BibitemOpen
  \bibfield  {author} {\bibinfo {author} {\bibfnamefont {M.~P.}\ \bibnamefont
  {M{\"u}ller}},\ }\href@noop {} {\bibfield  {journal} {\bibinfo  {journal}
  {Physical Review X}\ }\textbf {\bibinfo {volume} {8}},\ \bibinfo {pages}
  {041051} (\bibinfo {year} {2018})}\BibitemShut {NoStop}%
\bibitem [{\citenamefont {Shiraishi}\ and\ \citenamefont
  {Sagawa}(2021)}]{shiraishi2021quantum}%
  \BibitemOpen
  \bibfield  {author} {\bibinfo {author} {\bibfnamefont {N.}~\bibnamefont
  {Shiraishi}}\ and\ \bibinfo {author} {\bibfnamefont {T.}~\bibnamefont
  {Sagawa}},\ }\href@noop {} {\bibfield  {journal} {\bibinfo  {journal}
  {Physical Review Letters}\ }\textbf {\bibinfo {volume} {126}},\ \bibinfo
  {pages} {150502} (\bibinfo {year} {2021})}\BibitemShut {NoStop}%
\bibitem [{\citenamefont {Wilming}\ \emph {et~al.}(2017)\citenamefont
  {Wilming}, \citenamefont {Gallego},\ and\ \citenamefont
  {Eisert}}]{wilming2017axiomatic}%
  \BibitemOpen
  \bibfield  {author} {\bibinfo {author} {\bibfnamefont {H.}~\bibnamefont
  {Wilming}}, \bibinfo {author} {\bibfnamefont {R.}~\bibnamefont {Gallego}}, \
  and\ \bibinfo {author} {\bibfnamefont {J.}~\bibnamefont {Eisert}},\
  }\href@noop {} {\bibfield  {journal} {\bibinfo  {journal} {Entropy}\ }\textbf
  {\bibinfo {volume} {19}},\ \bibinfo {pages} {241} (\bibinfo {year}
  {2017})}\BibitemShut {NoStop}%
\bibitem [{\citenamefont {Boes}\ \emph {et~al.}(2019)\citenamefont {Boes},
  \citenamefont {Eisert}, \citenamefont {Gallego}, \citenamefont {M{\"u}ller},\
  and\ \citenamefont {Wilming}}]{boes2019neumann}%
  \BibitemOpen
  \bibfield  {author} {\bibinfo {author} {\bibfnamefont {P.}~\bibnamefont
  {Boes}}, \bibinfo {author} {\bibfnamefont {J.}~\bibnamefont {Eisert}},
  \bibinfo {author} {\bibfnamefont {R.}~\bibnamefont {Gallego}}, \bibinfo
  {author} {\bibfnamefont {M.~P.}\ \bibnamefont {M{\"u}ller}}, \ and\ \bibinfo
  {author} {\bibfnamefont {H.}~\bibnamefont {Wilming}},\ }\href@noop {}
  {\bibfield  {journal} {\bibinfo  {journal} {Physical Review Letters}\
  }\textbf {\bibinfo {volume} {122}},\ \bibinfo {pages} {210402} (\bibinfo
  {year} {2019})}\BibitemShut {NoStop}%
\bibitem [{\citenamefont {Rethinasamy}\ and\ \citenamefont
  {Wilde}(2020)}]{rethinasamy2020relative}%
  \BibitemOpen
  \bibfield  {author} {\bibinfo {author} {\bibfnamefont {S.}~\bibnamefont
  {Rethinasamy}}\ and\ \bibinfo {author} {\bibfnamefont {M.~M.}\ \bibnamefont
  {Wilde}},\ }\href@noop {} {\bibfield  {journal} {\bibinfo  {journal}
  {Physical Review Research}\ }\textbf {\bibinfo {volume} {2}},\ \bibinfo
  {pages} {033455} (\bibinfo {year} {2020})}\BibitemShut {NoStop}%
\bibitem [{\citenamefont {Kondra}\ \emph {et~al.}(2021)\citenamefont {Kondra},
  \citenamefont {Datta},\ and\ \citenamefont
  {Streltsov}}]{kondra2021catalytic}%
  \BibitemOpen
  \bibfield  {author} {\bibinfo {author} {\bibfnamefont {T.~V.}\ \bibnamefont
  {Kondra}}, \bibinfo {author} {\bibfnamefont {C.}~\bibnamefont {Datta}}, \
  and\ \bibinfo {author} {\bibfnamefont {A.}~\bibnamefont {Streltsov}},\
  }\href@noop {} {\bibfield  {journal} {\bibinfo  {journal} {arXiv preprint
  arXiv:2102.11136}\ } (\bibinfo {year} {2021})}\BibitemShut {NoStop}%
\bibitem [{\citenamefont {Lie}\ and\ \citenamefont
  {Jeong}(2020)}]{lie2020randomness}%
  \BibitemOpen
  \bibfield  {author} {\bibinfo {author} {\bibfnamefont {S.~H.}\ \bibnamefont
  {Lie}}\ and\ \bibinfo {author} {\bibfnamefont {H.}~\bibnamefont {Jeong}},\
  }\href@noop {} {\bibfield  {journal} {\bibinfo  {journal} {Physical Review
  A}\ }\textbf {\bibinfo {volume} {101}},\ \bibinfo {pages} {052322} (\bibinfo
  {year} {2020})}\BibitemShut {NoStop}%
\bibitem [{\citenamefont {Lie}\ and\ \citenamefont
  {Jeong}(2021{\natexlab{a}})}]{lie2020uniform}%
  \BibitemOpen
  \bibfield  {author} {\bibinfo {author} {\bibfnamefont {S.~H.}\ \bibnamefont
  {Lie}}\ and\ \bibinfo {author} {\bibfnamefont {H.}~\bibnamefont {Jeong}},\
  }\href@noop {} {\bibfield  {journal} {\bibinfo  {journal} {Physical Review
  Research}\ }\textbf {\bibinfo {volume} {3}},\ \bibinfo {pages} {013218}
  (\bibinfo {year} {2021}{\natexlab{a}})}\BibitemShut {NoStop}%
\bibitem [{\citenamefont {Lie}\ and\ \citenamefont
  {Jeong}(2021{\natexlab{b}})}]{lie2021correlational}%
  \BibitemOpen
  \bibfield  {author} {\bibinfo {author} {\bibfnamefont {S.~H.}\ \bibnamefont
  {Lie}}\ and\ \bibinfo {author} {\bibfnamefont {H.}~\bibnamefont {Jeong}},\
  }\href@noop {} {\bibfield  {journal} {\bibinfo  {journal} {arXiv preprint
  arXiv:2104.00300}\ } (\bibinfo {year} {2021}{\natexlab{b}})}\BibitemShut
  {NoStop}%
\bibitem [{\citenamefont {Lie}\ and\ \citenamefont
  {Jeong}(2022)}]{lie2022delocalized}%
  \BibitemOpen
  \bibfield  {author} {\bibinfo {author} {\bibfnamefont {S.~H.}\ \bibnamefont
  {Lie}}\ and\ \bibinfo {author} {\bibfnamefont {H.}~\bibnamefont {Jeong}},\
  }\href@noop {} {\bibfield  {journal} {\bibinfo  {journal} {arXiv preprint
  arXiv:2206.11469}\ } (\bibinfo {year} {2022})}\BibitemShut {NoStop}%
\bibitem [{\citenamefont {Sparaciari}(2018)}]{sparaciari2018multi}%
  \BibitemOpen
  \bibfield  {author} {\bibinfo {author} {\bibfnamefont {C.}~\bibnamefont
  {Sparaciari}},\ }\emph {\bibinfo {title} {Multi-resource theories and
  applications to quantum thermodynamics}},\ \href@noop {} {Ph.D. thesis},\
  \bibinfo  {school} {UCL (University College London)} (\bibinfo {year}
  {2018})\BibitemShut {NoStop}%
\bibitem [{\citenamefont {Feng}\ \emph {et~al.}(2005)\citenamefont {Feng},
  \citenamefont {Duan},\ and\ \citenamefont {Ying}}]{feng2005catalyst}%
  \BibitemOpen
  \bibfield  {author} {\bibinfo {author} {\bibfnamefont {Y.}~\bibnamefont
  {Feng}}, \bibinfo {author} {\bibfnamefont {R.}~\bibnamefont {Duan}}, \ and\
  \bibinfo {author} {\bibfnamefont {M.}~\bibnamefont {Ying}},\ }\href@noop {}
  {\bibfield  {journal} {\bibinfo  {journal} {IEEE transactions on information
  theory}\ }\textbf {\bibinfo {volume} {51}},\ \bibinfo {pages} {1090}
  (\bibinfo {year} {2005})}\BibitemShut {NoStop}%
\bibitem [{\citenamefont {Fritz}(2017)}]{fritz2017resource}%
  \BibitemOpen
  \bibfield  {author} {\bibinfo {author} {\bibfnamefont {T.}~\bibnamefont
  {Fritz}},\ }\href@noop {} {\bibfield  {journal} {\bibinfo  {journal}
  {Mathematical Structures in Computer Science}\ }\textbf {\bibinfo {volume}
  {27}},\ \bibinfo {pages} {850} (\bibinfo {year} {2017})}\BibitemShut
  {NoStop}%
\bibitem [{\citenamefont {Wilde}\ and\ \citenamefont
  {Mizel}(2012)}]{wilde2012addressing}%
  \BibitemOpen
  \bibfield  {author} {\bibinfo {author} {\bibfnamefont {M.~M.}\ \bibnamefont
  {Wilde}}\ and\ \bibinfo {author} {\bibfnamefont {A.}~\bibnamefont {Mizel}},\
  }\href@noop {} {\bibfield  {journal} {\bibinfo  {journal} {Foundations of
  Physics}\ }\textbf {\bibinfo {volume} {42}},\ \bibinfo {pages} {256}
  (\bibinfo {year} {2012})}\BibitemShut {NoStop}%
\bibitem [{\citenamefont {Leggett}\ and\ \citenamefont
  {Garg}(1985)}]{leggett1985quantum}%
  \BibitemOpen
  \bibfield  {author} {\bibinfo {author} {\bibfnamefont {A.~J.}\ \bibnamefont
  {Leggett}}\ and\ \bibinfo {author} {\bibfnamefont {A.}~\bibnamefont {Garg}},\
  }\href@noop {} {\bibfield  {journal} {\bibinfo  {journal} {Physical Review
  Letters}\ }\textbf {\bibinfo {volume} {54}},\ \bibinfo {pages} {857}
  (\bibinfo {year} {1985})}\BibitemShut {NoStop}%
\bibitem [{\citenamefont {Clemente}\ and\ \citenamefont
  {Kofler}(2016)}]{clemente2016no}%
  \BibitemOpen
  \bibfield  {author} {\bibinfo {author} {\bibfnamefont {L.}~\bibnamefont
  {Clemente}}\ and\ \bibinfo {author} {\bibfnamefont {J.}~\bibnamefont
  {Kofler}},\ }\href@noop {} {\bibfield  {journal} {\bibinfo  {journal}
  {Physical Review Letters}\ }\textbf {\bibinfo {volume} {116}},\ \bibinfo
  {pages} {150401} (\bibinfo {year} {2016})}\BibitemShut {NoStop}%
\bibitem [{\citenamefont {Pan}(2020)}]{pan2020interference}%
  \BibitemOpen
  \bibfield  {author} {\bibinfo {author} {\bibfnamefont {A.}~\bibnamefont
  {Pan}},\ }\href@noop {} {\bibfield  {journal} {\bibinfo  {journal} {Physical
  Review A}\ }\textbf {\bibinfo {volume} {102}},\ \bibinfo {pages} {032206}
  (\bibinfo {year} {2020})}\BibitemShut {NoStop}%
\bibitem [{\citenamefont {Lie}\ and\ \citenamefont
  {Jeong}(2023)}]{lie2023faithfulness}%
  \BibitemOpen
  \bibfield  {author} {\bibinfo {author} {\bibfnamefont {S.~H.}\ \bibnamefont
  {Lie}}\ and\ \bibinfo {author} {\bibfnamefont {H.}~\bibnamefont {Jeong}},\
  }\href@noop {} {\bibfield  {journal} {\bibinfo  {journal} {Physical Review
  Letters}\ }\textbf {\bibinfo {volume} {130}},\ \bibinfo {pages} {020802}
  (\bibinfo {year} {2023})}\BibitemShut {NoStop}%
\bibitem [{\citenamefont {Wolf}(2012)}]{wolf2012quantum}%
  \BibitemOpen
  \bibfield  {author} {\bibinfo {author} {\bibfnamefont {M.~M.}\ \bibnamefont
  {Wolf}},\ }\href@noop {} {\bibfield  {journal} {\bibinfo  {journal} {Lecture
  notes available at http://www-m5. ma. tum. de/foswiki/pub M}\ }\textbf
  {\bibinfo {volume} {5}} (\bibinfo {year} {2012})}\BibitemShut {NoStop}%
\bibitem [{\citenamefont {Artin}(1927)}]{artin1927theorie}%
  \BibitemOpen
  \bibfield  {author} {\bibinfo {author} {\bibfnamefont {E.}~\bibnamefont
  {Artin}},\ }in\ \href@noop {} {\emph {\bibinfo {booktitle} {Abhandlungen aus
  dem Mathematischen Seminar der Universit{\"a}t Hamburg}}},\ Vol.~\bibinfo
  {volume} {5}\ (\bibinfo {organization} {Springer},\ \bibinfo {year} {1927})\
  pp.\ \bibinfo {pages} {251--260}\BibitemShut {NoStop}%
\bibitem [{\citenamefont {Wedderburn}(1908)}]{wedderburn1908hypercomplex}%
  \BibitemOpen
  \bibfield  {author} {\bibinfo {author} {\bibfnamefont {J.}~\bibnamefont
  {Wedderburn}},\ }\href@noop {} {\bibfield  {journal} {\bibinfo  {journal}
  {Proceedings of the London Mathematical Society}\ }\textbf {\bibinfo {volume}
  {2}},\ \bibinfo {pages} {77} (\bibinfo {year} {1908})}\BibitemShut {NoStop}%
\bibitem [{\citenamefont {Brun}(2002)}]{brun2002simple}%
  \BibitemOpen
  \bibfield  {author} {\bibinfo {author} {\bibfnamefont {T.~A.}\ \bibnamefont
  {Brun}},\ }\href@noop {} {\bibfield  {journal} {\bibinfo  {journal} {American
  Journal of Physics}\ }\textbf {\bibinfo {volume} {70}},\ \bibinfo {pages}
  {719} (\bibinfo {year} {2002})}\BibitemShut {NoStop}%
\bibitem [{\citenamefont {Gudder}(2005)}]{gudder2005non}%
  \BibitemOpen
  \bibfield  {author} {\bibinfo {author} {\bibfnamefont {S.}~\bibnamefont
  {Gudder}},\ }\href@noop {} {\bibfield  {journal} {\bibinfo  {journal} {Fuzzy
  sets and systems}\ }\textbf {\bibinfo {volume} {155}},\ \bibinfo {pages} {18}
  (\bibinfo {year} {2005})}\BibitemShut {NoStop}%
\bibitem [{\citenamefont {Winter}(1999)}]{winter1999coding}%
  \BibitemOpen
  \bibfield  {author} {\bibinfo {author} {\bibfnamefont {A.}~\bibnamefont
  {Winter}},\ }\href@noop {} {\bibfield  {journal} {\bibinfo  {journal} {IEEE
  Transactions on Information Theory}\ }\textbf {\bibinfo {volume} {45}},\
  \bibinfo {pages} {2481} (\bibinfo {year} {1999})}\BibitemShut {NoStop}%
\bibitem [{\citenamefont {Streltsov}\ \emph {et~al.}(2011)\citenamefont
  {Streltsov}, \citenamefont {Kampermann},\ and\ \citenamefont
  {Bru{\ss}}}]{streltsov2011linking}%
  \BibitemOpen
  \bibfield  {author} {\bibinfo {author} {\bibfnamefont {A.}~\bibnamefont
  {Streltsov}}, \bibinfo {author} {\bibfnamefont {H.}~\bibnamefont
  {Kampermann}}, \ and\ \bibinfo {author} {\bibfnamefont {D.}~\bibnamefont
  {Bru{\ss}}},\ }\href@noop {} {\bibfield  {journal} {\bibinfo  {journal}
  {Physical Review Letters}\ }\textbf {\bibinfo {volume} {106}},\ \bibinfo
  {pages} {160401} (\bibinfo {year} {2011})}\BibitemShut {NoStop}%
\bibitem [{\citenamefont {Bennett}\ and\ \citenamefont
  {Brassard}(2014)}]{bennett2014quantum}%
  \BibitemOpen
  \bibfield  {author} {\bibinfo {author} {\bibfnamefont {C.~H.}\ \bibnamefont
  {Bennett}}\ and\ \bibinfo {author} {\bibfnamefont {G.}~\bibnamefont
  {Brassard}},\ }\href@noop {} {\bibfield  {journal} {\bibinfo  {journal}
  {Theoretical Computer Science}\ }\textbf {\bibinfo {volume} {560}},\ \bibinfo
  {pages} {7} (\bibinfo {year} {2014})}\BibitemShut {NoStop}%
\bibitem [{\citenamefont {Wilde}(2013)}]{wilde2013quantum}%
  \BibitemOpen
  \bibfield  {author} {\bibinfo {author} {\bibfnamefont {M.~M.}\ \bibnamefont
  {Wilde}},\ }\href@noop {} {\emph {\bibinfo {title} {Quantum information
  theory}}}\ (\bibinfo  {publisher} {Cambridge University Press},\ \bibinfo
  {year} {2013})\BibitemShut {NoStop}%
\bibitem [{\citenamefont {Nayak}\ and\ \citenamefont
  {Sen}(2007)}]{nayak2007invertible}%
  \BibitemOpen
  \bibfield  {author} {\bibinfo {author} {\bibfnamefont {A.}~\bibnamefont
  {Nayak}}\ and\ \bibinfo {author} {\bibfnamefont {P.}~\bibnamefont {Sen}},\
  }\href@noop {} {\bibfield  {journal} {\bibinfo  {journal} {Quantum
  Information \& Computation}\ }\textbf {\bibinfo {volume} {7}},\ \bibinfo
  {pages} {103} (\bibinfo {year} {2007})}\BibitemShut {NoStop}%
\bibitem [{\citenamefont {Koashi}\ and\ \citenamefont
  {Imoto}(2002)}]{koashi2002operations}%
  \BibitemOpen
  \bibfield  {author} {\bibinfo {author} {\bibfnamefont {M.}~\bibnamefont
  {Koashi}}\ and\ \bibinfo {author} {\bibfnamefont {N.}~\bibnamefont {Imoto}},\
  }\href@noop {} {\bibfield  {journal} {\bibinfo  {journal} {Physical Review
  A}\ }\textbf {\bibinfo {volume} {66}},\ \bibinfo {pages} {022318} (\bibinfo
  {year} {2002})}\BibitemShut {NoStop}%
\bibitem [{\citenamefont {Hayden}\ \emph {et~al.}(2004)\citenamefont {Hayden},
  \citenamefont {Jozsa}, \citenamefont {Petz},\ and\ \citenamefont
  {Winter}}]{hayden2004structure}%
  \BibitemOpen
  \bibfield  {author} {\bibinfo {author} {\bibfnamefont {P.}~\bibnamefont
  {Hayden}}, \bibinfo {author} {\bibfnamefont {R.}~\bibnamefont {Jozsa}},
  \bibinfo {author} {\bibfnamefont {D.}~\bibnamefont {Petz}}, \ and\ \bibinfo
  {author} {\bibfnamefont {A.}~\bibnamefont {Winter}},\ }\href@noop {}
  {\bibfield  {journal} {\bibinfo  {journal} {Communications in mathematical
  physics}\ }\textbf {\bibinfo {volume} {246}},\ \bibinfo {pages} {359}
  (\bibinfo {year} {2004})}\BibitemShut {NoStop}%
\bibitem [{\citenamefont {Horodecki}\ \emph {et~al.}(2009)\citenamefont
  {Horodecki}, \citenamefont {Horodecki}, \citenamefont {Horodecki},\ and\
  \citenamefont {Horodecki}}]{horodecki2009quantum}%
  \BibitemOpen
  \bibfield  {author} {\bibinfo {author} {\bibfnamefont {R.}~\bibnamefont
  {Horodecki}}, \bibinfo {author} {\bibfnamefont {P.}~\bibnamefont
  {Horodecki}}, \bibinfo {author} {\bibfnamefont {M.}~\bibnamefont
  {Horodecki}}, \ and\ \bibinfo {author} {\bibfnamefont {K.}~\bibnamefont
  {Horodecki}},\ }\href@noop {} {\bibfield  {journal} {\bibinfo  {journal}
  {Reviews of modern physics}\ }\textbf {\bibinfo {volume} {81}},\ \bibinfo
  {pages} {865} (\bibinfo {year} {2009})}\BibitemShut {NoStop}%
\bibitem [{\citenamefont {Streltsov}\ \emph {et~al.}(2017)\citenamefont
  {Streltsov}, \citenamefont {Adesso},\ and\ \citenamefont
  {Plenio}}]{streltsov2017colloquium}%
  \BibitemOpen
  \bibfield  {author} {\bibinfo {author} {\bibfnamefont {A.}~\bibnamefont
  {Streltsov}}, \bibinfo {author} {\bibfnamefont {G.}~\bibnamefont {Adesso}}, \
  and\ \bibinfo {author} {\bibfnamefont {M.~B.}\ \bibnamefont {Plenio}},\
  }\href@noop {} {\bibfield  {journal} {\bibinfo  {journal} {Reviews of Modern
  Physics}\ }\textbf {\bibinfo {volume} {89}},\ \bibinfo {pages} {041003}
  (\bibinfo {year} {2017})}\BibitemShut {NoStop}%
\bibitem [{\citenamefont {Wu}\ \emph {et~al.}(2021)\citenamefont {Wu},
  \citenamefont {Kondra}, \citenamefont {Rana}, \citenamefont {Scandolo},
  \citenamefont {Xiang}, \citenamefont {Li}, \citenamefont {Guo},\ and\
  \citenamefont {Streltsov}}]{wu2021resource}%
  \BibitemOpen
  \bibfield  {author} {\bibinfo {author} {\bibfnamefont {K.-D.}\ \bibnamefont
  {Wu}}, \bibinfo {author} {\bibfnamefont {T.~V.}\ \bibnamefont {Kondra}},
  \bibinfo {author} {\bibfnamefont {S.}~\bibnamefont {Rana}}, \bibinfo {author}
  {\bibfnamefont {C.~M.}\ \bibnamefont {Scandolo}}, \bibinfo {author}
  {\bibfnamefont {G.-Y.}\ \bibnamefont {Xiang}}, \bibinfo {author}
  {\bibfnamefont {C.-F.}\ \bibnamefont {Li}}, \bibinfo {author} {\bibfnamefont
  {G.-C.}\ \bibnamefont {Guo}}, \ and\ \bibinfo {author} {\bibfnamefont
  {A.}~\bibnamefont {Streltsov}},\ }\href@noop {} {\bibfield  {journal}
  {\bibinfo  {journal} {Physical Review A}\ }\textbf {\bibinfo {volume}
  {103}},\ \bibinfo {pages} {032401} (\bibinfo {year} {2021})}\BibitemShut
  {NoStop}%
\bibitem [{\citenamefont {Nicholson}(1993)}]{nicholson1993short}%
  \BibitemOpen
  \bibfield  {author} {\bibinfo {author} {\bibfnamefont {W.~K.}\ \bibnamefont
  {Nicholson}},\ }\href@noop {} {\bibfield  {journal} {\bibinfo  {journal} {New
  Zealand J. Math}\ }\textbf {\bibinfo {volume} {22}},\ \bibinfo {pages} {83}
  (\bibinfo {year} {1993})}\BibitemShut {NoStop}%
\bibitem [{\citenamefont {Bre{\v{s}}ar}(2010)}]{brevsar2010elementary}%
  \BibitemOpen
  \bibfield  {author} {\bibinfo {author} {\bibfnamefont {M.}~\bibnamefont
  {Bre{\v{s}}ar}},\ }\href@noop {} {\bibfield  {journal} {\bibinfo  {journal}
  {Expositiones Mathematicae}\ }\textbf {\bibinfo {volume} {28}},\ \bibinfo
  {pages} {79} (\bibinfo {year} {2010})}\BibitemShut {NoStop}%
\bibitem [{\citenamefont {Burgarth}\ \emph {et~al.}(2013)\citenamefont
  {Burgarth}, \citenamefont {Chiribella}, \citenamefont {Giovannetti},
  \citenamefont {Perinotti},\ and\ \citenamefont
  {Yuasa}}]{burgarth2013ergodic}%
  \BibitemOpen
  \bibfield  {author} {\bibinfo {author} {\bibfnamefont {D.}~\bibnamefont
  {Burgarth}}, \bibinfo {author} {\bibfnamefont {G.}~\bibnamefont
  {Chiribella}}, \bibinfo {author} {\bibfnamefont {V.}~\bibnamefont
  {Giovannetti}}, \bibinfo {author} {\bibfnamefont {P.}~\bibnamefont
  {Perinotti}}, \ and\ \bibinfo {author} {\bibfnamefont {K.}~\bibnamefont
  {Yuasa}},\ }\href@noop {} {\bibfield  {journal} {\bibinfo  {journal} {New
  Journal of Physics}\ }\textbf {\bibinfo {volume} {15}},\ \bibinfo {pages}
  {073045} (\bibinfo {year} {2013})}\BibitemShut {NoStop}%
\bibitem [{\citenamefont {Schr{\"o}dinger}(1935)}]{schrodinger1935discussion}%
  \BibitemOpen
  \bibfield  {author} {\bibinfo {author} {\bibfnamefont {E.}~\bibnamefont
  {Schr{\"o}dinger}},\ }in\ \href@noop {} {\emph {\bibinfo {booktitle}
  {Mathematical Proceedings of the Cambridge Philosophical Society}}},\
  Vol.~\bibinfo {volume} {31}\ (\bibinfo {organization} {Cambridge University
  Press},\ \bibinfo {year} {1935})\ pp.\ \bibinfo {pages}
  {555--563}\BibitemShut {NoStop}%
\bibitem [{\citenamefont {Hughston}\ \emph {et~al.}(1993)\citenamefont
  {Hughston}, \citenamefont {Jozsa},\ and\ \citenamefont
  {Wootters}}]{hughston1993complete}%
  \BibitemOpen
  \bibfield  {author} {\bibinfo {author} {\bibfnamefont {L.~P.}\ \bibnamefont
  {Hughston}}, \bibinfo {author} {\bibfnamefont {R.}~\bibnamefont {Jozsa}}, \
  and\ \bibinfo {author} {\bibfnamefont {W.~K.}\ \bibnamefont {Wootters}},\
  }\href@noop {} {\bibfield  {journal} {\bibinfo  {journal} {Physics Letters
  A}\ }\textbf {\bibinfo {volume} {183}},\ \bibinfo {pages} {14} (\bibinfo
  {year} {1993})}\BibitemShut {NoStop}%
\bibitem [{Note1()}]{Note1}%
  \BibitemOpen
  \bibinfo {note} {This is also why the probabilities in Eq.~\protect \textup
  {\hbox {\mathsurround \z@ \protect \normalfont (\ignorespaces \ref
  {eqn:PCdec}\unskip \@@italiccorr )}} were ommitted in Eq.~\protect \textup
  {\hbox {\mathsurround \z@ \protect \normalfont (\ignorespaces \ref
  {eqn:qcdec}\unskip \@@italiccorr )}}, since w.l.o.g. they can be absorbed
  into $\protect \mathcal {T}_{C_i^R}$.}\BibitemShut {Stop}%
\bibitem [{\citenamefont {Takesaki}\ \emph {et~al.}(1979)\citenamefont
  {Takesaki} \emph {et~al.}}]{takesaki2003theory}%
  \BibitemOpen
  \bibfield  {author} {\bibinfo {author} {\bibfnamefont {M.}~\bibnamefont
  {Takesaki}} \emph {et~al.},\ }\href@noop {} {\emph {\bibinfo {title} {Theory
  of operator algebras I}}}\ (\bibinfo  {publisher} {Springer New York, NY},\
  \bibinfo {year} {1979})\BibitemShut {NoStop}%
\end{thebibliography}%

\end{document}